\documentclass{ieeeaccess}
\usepackage{amsmath,amsfonts}
\usepackage{algorithmic}
\usepackage{algorithm}
\usepackage{array}
\usepackage{textcomp}
\usepackage{stfloats}
\usepackage{url}
\usepackage{verbatim}
\usepackage{graphicx}
\usepackage{cite}
\usepackage{braket}
\usepackage{enumerate}
\usepackage{balance}
\usepackage{subfloat}
\usepackage{subfig}
\usepackage{breakurl}
\graphicspath{ {./images/} }

\def\BibTeX{{\rm B\kern-.05em{\sc i\kern-.025em b}\kern-.08em
    T\kern-.1667em\lower.7ex\hbox{E}\kern-.125emX}}

\begin{document}
\title{SPARQ: Efficient Entanglement Distribution and Routing in Space-Air-Ground Quantum Networks}

\author{\uppercase{Mohamed Shaban}\authorrefmark{1,2},
\uppercase{Muhammad Ismail}\authorrefmark{1}, \IEEEmembership{Senior Member, IEEE},
\uppercase{and Walid Saad}\authorrefmark{3}, \IEEEmembership{Fellow, IEEE}}

\address[1]{Cybersecurity Education, Research, and Outreach Center (CEROC) and Computer Science Department, Tennessee Tech University, TN, USA (e-mail: mmibrahims42@tntech.edu and mismail@tntech.edu)}
\address[2]{Department of Mathematics, Faculty of Education, Alexandria University, Egypt}
\address[3]{Bradley Department of Electrical and Computer Engineering, Virginia Tech, Blacksburg, VA 24061 USA (e-mail: walids@vt.edu)}

\tfootnote{This material is based on work supported by the National Science Foundation under Award No. 2210251.}


\corresp{Corresponding author: Mohamed Shaban (email: mmibrahims42@tntech.edu).}

\begin{abstract}
In this paper, a space-air-ground quantum (SPARQ) network is developed as a means for providing a seamless on-demand entanglement distribution. The node mobility in SPARQ poses significant challenges to entanglement routing. Existing quantum routing algorithms focus on stationary ground nodes and utilize link distance as an optimality metric, which is unrealistic for dynamic systems like SPARQ. Moreover, in contrast to the prior art that assumes homogeneous nodes, SPARQ encompasses heterogeneous nodes with different functionalities further complicates the entanglement distribution. To solve the entanglement routing problem, a deep reinforcement learning (RL) framework is proposed and trained using deep Q-network (DQN) on multiple graphs of SPARQ to account for the network dynamics. Subsequently, an entanglement distribution policy, third-party entanglement distribution (TPED), is proposed to establish entanglement between communication parties. A realistic quantum network simulator is designed for performance evaluation. Simulation results show that the TPED policy improves entanglement fidelity by $3\%$ and reduces memory consumption by $50\%$ compared with benchmark. The results also show that the proposed DQN algorithm improves the number of resolved teleportation requests by $39\%$ compared with shortest path baseline and the entanglement fidelity by $2\%$ compared with an RL algorithm that is based on long short-term memory (LSTM). It also improved entanglement fidelity by $6\%$ and $9\%$ compared with two state-of-the-art benchmarks. Moreover, the entanglement fidelity is improved by $15\%$ compared with DQN trained on a snapshot of SPARQ. Additionally, SPARQ enhances the average entanglement fidelity by $23.5\%$ compared with existing networks spanning only space and ground layers.

\end{abstract}


\titlepgskip=-15pt

\maketitle

\section{Introduction}
\PARstart{Q}{uantum} information science can potentially revolutionize various disciplines by addressing major challenges that are beyond the capabilities of classical computers. For instance, quantum computing will require a major re-evaluation of security measures because current security algorithms can be easily compromised by quantum computers. Although large quantum computers could solve certain computational problems substantially faster than classical computers \cite{138}, current quantum computers are still limited in resources to outperform their classical counterparts in practical applications. The development of large-scale quantum computers equipped with a sufficient number of qubits remains an open technological challenge that may be difficult to achieve in the foreseeable future. 

However, the potential of quantum computing can still be achieved through distributed quantum computing paradigms in which numerous quantum computers collaborate to tackle a single problem. In distributed quantum computing, the division of qubits across multiple quantum computers can present challenges, particularly when multi-qubit gates, such as the CNOT gate, are involved. Since the control and target qubits may reside on different computers, they must be physically co-located to perform the operation. To overcome this limitation, one of the techniques used is to teleport either the control or target qubit to the other system, where the gate operation can be performed, and then return it back to its original system. However, executing a large circuit in distributed quantum computing often requires frequent teleportation of qubits between systems. This process of teleporting qubits back and forth can lead to a reduction in computation speed, throughput, and increased latency if it is not carried out over an efficient network architecture. In this paradigm, quantum entanglement plays a pivotal role \cite{146}, serving as the fundamental resource for various quantum communication protocols, such as teleportation and superdense coding. In order to establish entanglement between two communication parties, entangled pairs are generated and transmitted through quantum channels. These channels suffer from losses that degrade the quality of the transmitted entangled photons. To overcome this limitation, several quantum repeaters can be used to perform entanglement swapping and directly entangle the source and destination. Subsequently, to enable distributed quantum computing, a resilient quantum network that enables seamless and on-demand entanglement distribution is required. Consequently, the following key open questions need to be answered: (a) What is the practical architecture for a near-term quantum network in terms of the type of adopted links and node functionality (sources of entanglement and repeaters)? (b) How to successfully establish an end-to-end path to create entanglement between two nodes, hence, supporting quantum teleportation and superdense coding? The establishment of such an end-to-end path can be performed with quantum routing.

\subsection{Related works}
\noindent{\textbf{Network Architecture:}}
Existing research on quantum-enabled space-air-ground networks is limited to quantum key distribution (QKD) services and does not develop a practical network architecture. The work in \cite{45} and \cite{46} presented quantum-enabled space-air-ground integrated networks that replace classical nodes with quantum nodes without covering the functionalities of such nodes and the specifications of the communication links. Furthermore, current space-ground projects such as China’s Micius \cite{102} and EuroQCI \cite{101} are not sufficient to achieve a fully functional quantum Internet. For instance, Micius focuses only on QKD services. Although EuroQCI aims to provide QKD and quantum communication to member states, it heavily relies on satellites for communication. However, satellite services are degraded during bad weather conditions and may not always be available hence leading to intermittent connectivity. In addition, the considerable length of satellite links poses a potential risk of degrading link quality. Therefore, the EuroQCI architecture lacks the capability to facilitate on-demand entanglement distribution. Hence, the network architecture question is still open. 

\vspace{1mm}

\noindent{\textbf{Entanglement Routing:}} Current research on quantum routing primarily focuses on ground-based networks. For instance, the works in \cite{26} and \cite{20} proposed quantum routing algorithms based on the expected end-to-end entanglement rate. Also, the authors in \cite{27} proposed a routing algorithm for grid quantum networks that is based on link distance. The work in \cite{24} investigated the impact of quantum routing algorithms on the key rate of QKD in grid quantum networks. In \cite{25}, entanglement was used to add quantum features to classical networks, and a shortest distance-based quantum routing algorithm was proposed. The work in \cite{23} developed a shortest path quantum routing algorithm for handling multiple requests simultaneously. Moreover, two neural networks are trained in \cite{21} to maximize the number of served requests within the network. The authors in \cite{22} proposed a routing algorithm that adopts end-to-end fidelity as a routing metric. Additionally, the works in \cite{111,112,113,114,115} used shortest path algorithms to solve the quantum routing problem. 
The authors in \cite{photonics11030268} developed a quantum routing algorithm for space-ground networks, but it is specifically designed for QKD services. Consequently, their algorithm may not effectively handle teleportation requests, as it prioritizes secret key rates as the optimality metric. Several factors restrict the applicability of existing quantum routing algorithms \cite{20,21,22,23,24,25,26,27,111,112,113,114,115,photonics11030268} in SPARQ: (a) utilizing the link distance to be the routing metric is impractical in SPARQ due to the significant variation in path lengths spanning three layers, (b) existing algorithms require intensive computations, restricting their scalability in large networks, (c) existing algorithm requires calculating the optimal path for each communication request, leading to delays that impede the on-demand entanglement distribution capabilities of SPARQ, and (d) existing algorithms are designed for static networks, limiting their applicability in SPARQ.

\vspace{1mm}

\noindent{\textbf{Entanglement Distribution:}} 
The work in \cite{jones2013highspeed} proposed a scheme for entanglement generation called Midpoint-source (MPS). This scheme utilizes a photon source positioned at the midpoint of an optical channel to generate entanglement between quantum dots. The receivers subsequently establish local entanglement between the communication qubit and the emitted photons with the help of Bell state measurement. The scheme is designed to be less susceptible to losses and to achieve the highest entanglement rate by repeatedly attempting to produce Bell states. However, MPS operates under the assumption of direct communication between parties, without involving intermediate repeaters. Furthermore, it primarily focuses on hardware aspects and lacks a specific policy for nodes to follow.

Several efforts have been made to design entanglement distribution policies. For instance, the work in \cite{126} proposed a genetic algorithm to discover an optimal policy for entanglement generation and distribution. Despite its potential optimality, this policy is computationally intensive and is susceptible to delays. Alternatively, \cite{127} employed a reinforcement learning (RL) agent to handle the task of entanglement distribution. However, this method is limited to a linear repeater chain comprising a maximum of five nodes. Additionally, the work in \cite{128} used dynamic programming and RL to propose an entanglement distribution policy. The works in \cite{129} and \cite{130} proposed policies for entanglement distribution that are based on Markov decision processes. Moreover, the work in \cite{137} tackled the entanglement distribution challenge by formulating it as a mixed-integer nonlinear programming optimization problem, aiming to optimize the entanglement generation rate. However, it is limited to small-scale quantum networks. The work in \cite{140} utilized exhaustive search techniques to establish entanglement between two nodes while meeting the specified quality of service requirement. This is achieved through the manipulation of various parameters, including the distance between adjacent nodes and the number of entanglement distillations conducted. However, using the exhaustive search can be computationally expensive for a large search space. Despite these research efforts, a common limitation among existing studies \cite{126,127,128,129,130,137} is the assumption of uniform functionalities across network nodes which restricts their applicability in SPARQ. This limitation arises from the fact that not all nodes within SPARQ possess the capability to generate entanglement. While some existing work considers a heterogeneous network \cite{137}, it typically assumes heterogeneity in network specifications, such as varying distances, diverse applications, and various quality-of-service requirements. In contrast, a more realistic assumption of heterogeneity would be to consider the functionalities and capabilities of the nodes. Additionally, determining the optimal strategy for each communication request introduces additional delays that scale with the number of nodes, restricting their scalability in large networks like SPARQ. Thus, there is a need for an entanglement distribution policy that establishes entanglement between communication parties through a set of intermediate nodes. This policy should specify the functionality of each intermediate node as a source of entanglement and/or repeater in a way that minimizes the number of entanglement swapping operations, thereby improving the end-to-end entanglement quality and reducing memory consumption by intermediate repeaters. Also, this policy should consider network limitations such as the incapability of ground nodes to serve as sources of entanglement.

\vspace{1mm}

\noindent{\textbf{Quantum Network Simulator:}} Existing quantum network simulators, such as QuNetSim \cite{41}, NetSquid \cite{117}, QDNS \cite{118}, SQUANCH \cite{120}, SeQUeNCe \cite{121}, SimulaQron \cite{122}, and SimQN \cite{10024900} are primarily focused on ground networks and terrestrial communications. However, to simulate large-scale space-air-ground quantum networks like SPARQ, there is a need for a quantum network simulator that considers satellites and their movements, high altitude platforms (HAPs) and their locations, and various communication channels such as fiber optical and free-space optical channels.

\subsection{Contributions}
The main contribution of the paper is proposing a space-air-ground quantum (SPARQ) network, which integrates satellites, HAPs, and ground end-users to distribute entanglements and create end-to-end on-demand entangled paths. This approach introduces an air layer, shortening the links within the entanglement path and enhancing the overall quality of communication links. Hence, SPARQ has the potential to harness the global quantum Internet, enabling applications such as distributed quantum computing, quantum communications, quantum sensing, and quantum security. SPARQ has the potential to address the limitations of the prior works. To achieve this overarching goal, our key contributions include:

\begin{itemize}
    \item We propose the SPARQ network architecture, which spans space, air, and ground layers. Furthermore, we specify node functionalities such as practical sources of entanglement, repeaters, and communication links. SPARQ facilitates quantum teleportation and superdense coding within the emerging quantum Internet, enabling the efficient realization of advanced network applications like distributed quantum computing. Simulation results demonstrate the superior performance of the proposed architecture compared to traditional satellite-terrestrial networks, exhibiting a remarkable improvement by $23.5\%$ in the average end-to-end entanglement fidelity.
    \item We propose a deep RL strategy to solve the entanglement routing problem within the SPARQ network. The deep RL agent is trained following a deep Q-network (DQN) policy. Unlike the traditional DQN policies, our proposed DQN agent is trained on the dynamic network topology of SPARQ, enabling it to capture comprehensive network graph states and develop an optimal policy for real-time entanglement routing. Simulation results showed that the proposed DQN algorithm achieves a $39\%$ improvement in the number of resolved teleportation requests compared to the shortest path baseline.
    Additionally, it improves the average entanglement fidelity by $15\%$ compared with another DQN algorithm trained on a snapshot of SPARQ. Moreover, the proposed DQN model achieves up to $9\%$ improvement in the average entanglement fidelity compared with state-of-the-art benchmark algorithms. Also, the simulation results show that the proposed DQN algorithm has minimal overhead compared to all considered baselines. \textcolor{black}{The proposed DQN algorithm also significantly reduces the time required to find the routing path, demonstrating a substantial improvement in speed compared to state-of-the-art algorithms.}

    \item We propose a new policy for on-demand entanglement distribution within SPARQ, called third-party entanglement distribution (TPED). This policy is designed to establish end-to-end entanglement between communication parties after determining the optimal routing path. The TPED policy assigns functionalities to each node in the routing path to act as sources of entanglement and/or repeaters in a way that minimizes the number of entanglement swapping operations, while also considering network node limitations such as the incapability of ground nodes to act as a source of entanglement. The goal of TPED is to enhance the quality of established end-to-end entanglement by reducing the number of entangled pairs consumed by intermediate repeaters. Simulation results demonstrate that TPED improved the average entanglement fidelity while reducing the memory consumption by $50\%$ compared with a benchmark. The benchmark approach involved establishing entanglement between adjacent nodes and subsequently performing entanglement swapping to achieve direct entanglement between the source and destination.
    
\end{itemize}

\textcolor{black}{We conducted a comprehensive evaluation of the proposed SPARQ architecture and algorithms by upgrading the QuNetSim simulator and integrating it with Ansys Systems Tool Kit (STK). This integration accounts for realistic space-air-ground node behaviors and models.} The rest of this paper is organized as follows. Section \ref{sec:Network Model} presents the proposed SPARQ network architecture and the channel models.   
Section \ref{sec:quantum_routing} defines the quantum routing problem within SPARQ and presents the proposed deep RL routing model for finding the optimal end-to-end entanglement path. Section \ref{sec:entanglement_distribution} proposes an on-demand entanglement distribution policy within SPARQ. Section \ref{sec:results} presents the SPARQ simulator which is used for performance evaluations, discusses the simulation setup, and presents the simulation results. Conclusions are drawn in Section \ref{sec:Conclusion}.

\section{SPARQ Architecture and Channel Model}
\label{sec:Network Model}

\subsection{Network Layers and Node Functionalities}
\begin{figure}[!t]
\centering
\includegraphics[width=3.5in]{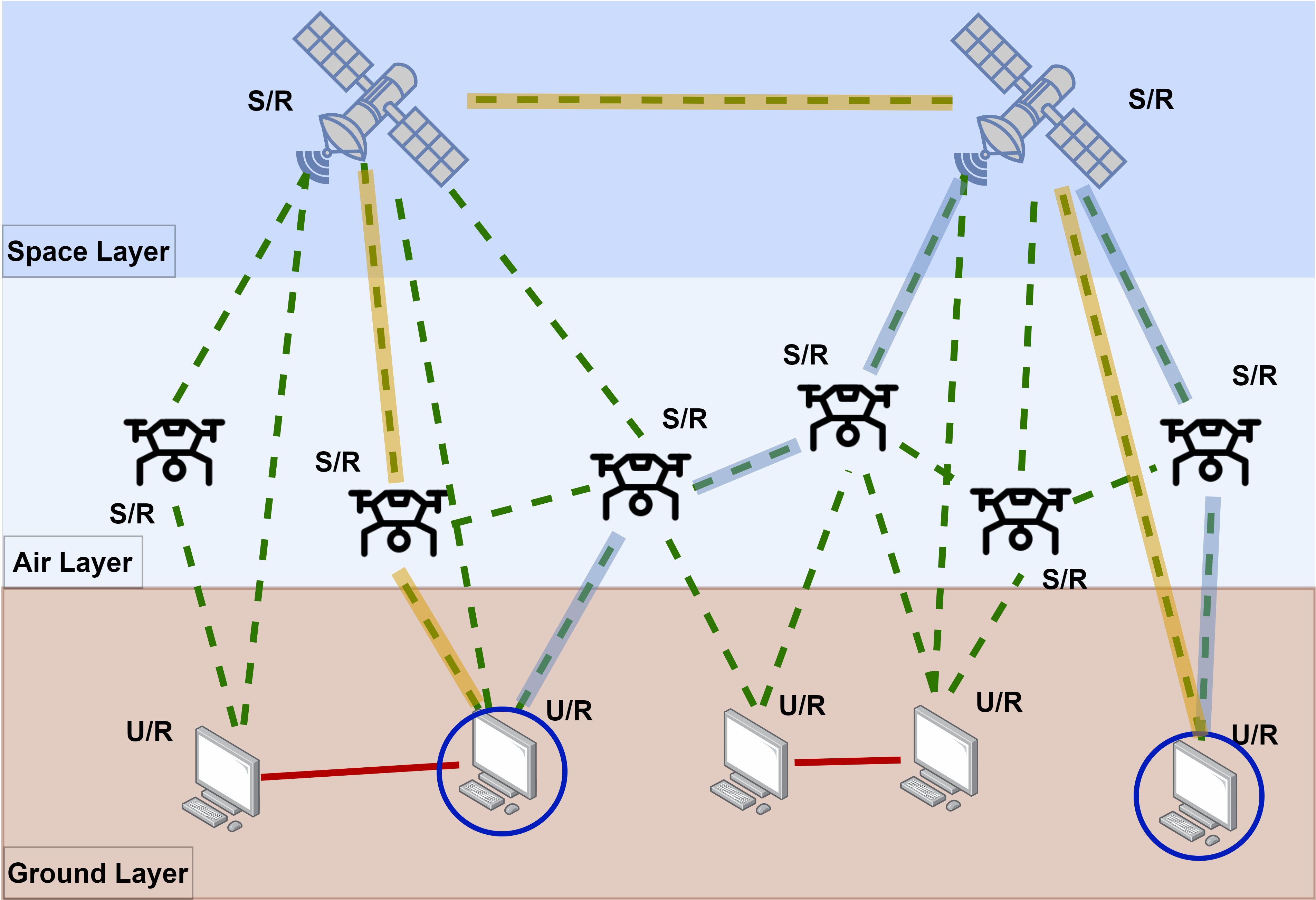}
\caption{Proposed architecture of SPARQ. FSO channels are shown with green dashed lines and optical fiber channels are shown with red solid lines. S, R, and U indicate a source of entanglement, repeater, and end-user, respectively.}
\label{fig:net}
\end{figure}

As shown in the architecture of Fig. \ref{fig:net}, the SPARQ network spans three layers as follows:
\begin{itemize}
    \item \textit{Space layer:} This layer consists of a satellite constellation equipped with quantum devices that can generate entangled states and manipulate qubits through measurements and processing. We consider that the satellites are positioned in a low earth orbit (LEO) at an altitude of $500$ km.
    
    \item \textit{Air layer:} This layer consists of swarms of aerial vehicles equipped with limited quantum capabilities, including a restricted number of qubits for generating entangled states and manipulating qubits through measurements and processing. The deployment of the air layer is motivated by the fact that satellite services may experience interruptions due to bad weather conditions and orbit limitations \cite{45}. Additionally, aerial vehicles can serve as relays between ground nodes and satellites, extending communication range and enhancing link quality. Consequently, this layer is deployed to address network demands when satellite services are unavailable and to bridge the communication gap between the space and ground layers, contributing to the development of a resilient network architecture. We use HAP drones at a height of $50$ km. 

    \item \textit{Ground layer:} This layer is composed of end-user quantum nodes with limited quantum capabilities, allowing them to measure and process qubits.
\end{itemize}

As mentioned earlier, it is necessary to distribute entangled pairs among end users in order to enable quantum communications within SPARQ. This process involves the creation of entangled pairs and their transmission through quantum channels. However, these channels are prone to losses and transmitted photons decay rapidly as the transmission distance increases. To address this challenge, multiple quantum repeaters can be adopted, allowing for entanglement swapping and the extension of the entanglement range. Consequently, the node functionalities in SPARQ can be summarized as follows:
\begin{itemize}
    \item \textit{Source of entanglement:} 
    These nodes are responsible for the generation of entangled pairs, which are the fuel for quantum networks. In SPARQ, satellites and HAPs are used as sources of entanglement.
    \item \textit{Repeaters:} These nodes are responsible for performing entanglement swapping to enhance the quality of transmitted photons over long distances. In SPARQ, satellites, HAPs, and ground end-users can act as repeaters. 
    \item \textit{End-users:} These are the ground nodes. Entanglement routing algorithms are employed to connect two communicating parties within the ground layer. The routing path makes use of intermediate nodes to establish an end-to-end entanglement connection between any two ground nodes.
\end{itemize}

\subsection{Communication Links}
Quantum communications can be realized through optical fiber channels or free space optical (FSO) channels, both technologies are adopted within SPARQ. Optical fiber offers certain advantages, such as establishing a reliable and uninterrupted connection between communicating parties. Moreover, photons transmitted over fiber links are typically unaffected by external factors such as background light or bad weather conditions. However, optical fiber technologies limit the transmission of photons to a few hundred kilometers due to issues such as polarization preservation, random scattering, and optical attenuation. Conversely, FSO technology outperforms fiber optical technology in terms of losses because of the low atmospheric absorption \cite{50}. In addition, most of the transmission path for satellite-based FSO communication remains unaffected by either absorption or turbulence, as the atmospheric width is less than $10$ km \cite{50}. It is worth noting that FSO channels are used for real projects such as Micius \cite{102} and are widely employed for QKD in research \cite{45,46,141,142,143,144,145}. Therefore, FSO channels are adopted for communication between satellites, HAPs, and between satellites or HAPs and end-users. On the other hand, fiber optic channels are employed for communication between ground end-users. 

\subsection{Channel Models}
\label{sec:channel_model}
We consider two channels in our SPARQ system: fiber optic and FSO channels. In each channel, we consider transmissivity to address optical losses occurring during communication. Thus, while transmitting quantum photons over a communication channel, their decay is determined by the calculated transmissivity for that specific channel.
\subsubsection{Fiber optical channels}
For an attenuation coefficient $\alpha$ and transmission distance $l$, the transmissivity of fiber optical channels given by \cite{106}:
\begin{equation}
    \label{eq:fiber_t}
    \eta=e^{-\alpha l}.
\end{equation}

\subsubsection{FSO channels}
\label{sec:fso}
The transmissivity of the FSO channel considers the total optical loss that may occur in strong turbulence weather. Specifically, the transmissivity $\eta$ considers three optical transmissivities as follows \cite{107}:
\begin{equation}
    \label{eq:fso_t}
    \eta= \eta_{\mbox{\tiny\itshape lt}}\eta_{\mbox{\tiny\itshape eff}}\eta_{\mbox{\tiny\itshape atm}},
\end{equation}
where $\eta_{\mbox{\tiny\itshape lt}}$, $\eta_{\mbox{\tiny\itshape eff}}$, and $\eta_{\mbox{\tiny\itshape atm}}$ present the transmissivity based on turbulence, receiver efficiency, and atmospheric loss, respectively.  Herein, $\eta_{\mbox{\tiny\itshape atm}}$ is modeled by the Beer-Lambert equation \cite{107}:
\begin{equation}
    \eta_{\mbox{\tiny\itshape atm}}=e^{-\alpha(\lambda,h_0)z}, \alpha(\lambda,h_0)=\alpha_0(\lambda)e^{-\frac{h_0}{6600}},
\end{equation}
where $z$ is the propagation path length, $\lambda$ is the carrier wavelength, $h_0$ represents the altitude in meters, and $\alpha_0(\lambda)$ represents the extinction factor at sea level.

Furthermore, the transmissivity based on turbulence $\eta_{\mbox{\tiny\itshape lt}}$ is given by \cite{107}:
\begin{equation}
    \eta_{\mbox{\tiny\itshape lt}}=1-e^\frac{-2a_{\mbox{\tiny\itshape R}}^2}{w_{\mbox{\tiny\itshape lt}}^2},           
\end{equation}
where $a_{\mbox{\tiny\itshape R}}$ is the aperture radius and $w_{\mbox{\tiny\itshape lt}}$ is the long-term beam waist, which can be derived from \cite{107}:
\begin{equation}
    w_{\mbox{\tiny\itshape lt}}=w_z\sqrt{1+\frac{4}{3}q\Lambda},          
\end{equation}
where $w_z$ is the diffraction beam waist, while the parameters $\Lambda$ and $q$ are given by \cite{107}:

\begin{equation}
    \label{eq:Lamda}
    \Lambda = \frac{2z}{kw_z^{2}},
\end{equation}

\begin{equation}
    \label{eq:q}
    q=0.74\sigma^2\sqrt[6]{M}, M=35.05\frac{z}{kl_0^2},
\end{equation}
and $l_0$ is the turbulence inner scale, $z$ is the propagation path length, $k=2\pi/\lambda$ is the wave number, and $\sigma^2$ is the Raytov variance, which is given by \cite{107}:
\begin{equation}
    \sigma^2=2.25C_n^2k^\frac{7}{6}z^\frac{11}{6},
\end{equation}
where $C_n^2$ is the refractive index structure that is used to characterize the atmospheric turbulence strength.

One characteristic of natural light is diffraction, a phenomenon that continuously expands the beam waist of a wave as it travels through free space. This phenomenon also leads to a constant increase in the radius of curvature of the propagating beam. We start with a Gaussian beam characterized by an initial field spot size of $w_0$, a carrier wavelength $\lambda$, and an initial radius of curvature $R_0$. As the beam propagates to a distance $z$ where a receiver is positioned, free-space diffraction results in an enlargement of the beam's spot size to \cite{107}:

\begin{equation}
    w_z^2=w_0^2 \left[ \left(1-\frac{z}{R_0} \right)^2+\left(\frac{z}{z_{\mbox{\tiny\itshape R}}}\right)^2 \right],          
\end{equation}
where $z_{\mbox{\tiny\itshape R}}=\pi w_0^2/ \lambda$ is the beam’s Rayleigh length. 

\subsubsection{Amplitude damping channel}
The amplitude damping channel is a concept that describes a type of quantum channel through which a quantum system experiences a reduction in its amplitude or damping of its state over time. This channel is used to model certain physical processes that can cause the loss of quantum information or the decay of quantum states. In quantum communication, the amplitude damping channel can be used to model the attenuation that can occur during transmission. We adopt the amplitude damping channel to degrade the entanglement quality based on the transmissivity $\eta$. Then, the amplitude damping channel can be represented by the Kraus operator as follows \cite{110,103,104}:

\begin{equation}
    \label{eq:kraus}
    \boldsymbol{K_0}=\begin{bmatrix}
        1 & 0 \\
        0 & \sqrt{\eta} 
    \end{bmatrix},
    \boldsymbol{K_1}=\begin{bmatrix}
        0 & \sqrt{1-\eta} \\
        0 & 0 
    \end{bmatrix}.
\end{equation}

The action of the amplitude damping channel on the density matrix $\boldsymbol{\rho}$ is given by:
\begin{equation}
    \label{eq:damping}
    \boldsymbol{\rho'}=\boldsymbol{K_0} \boldsymbol{\rho} \boldsymbol{K_0}^\dag+\boldsymbol{K_1} \boldsymbol{\rho} \boldsymbol{K_1}^\dag.
\end{equation}

\subsubsection{Entanglement Fidelity}
Entanglement fidelity ($F_e$) is a metric used to quantify the similarity between an ideal entangled state and the actual state produced in a quantum system \cite{Tonolini2014}. Here, we use entanglement fidelity in order to measure the quality of the final end-to-end entanglement created between the source and the destination after going through all the channel losses experienced by the intermediate nodes. Entanglement fidelity is given by \cite{136} and \cite{Tonolini2014} as follows:

\begin{equation}
    \label{eq:fidelity}
     F_e = \left( \text{Tr} \left( \sqrt{\sqrt{\boldsymbol{\rho'}} \ket{\psi}\bra{\psi} \sqrt{\boldsymbol{\rho'}}} \right) \right)^2,
\end{equation}
where $\boldsymbol{\rho'}$ is the density matrix of the entangled state after experiencing losses, as given by \eqref{eq:damping}, and $\ket{\psi}$ represents the ideal entangled state. Here, $\ket{\psi}$ is the maximally entangled Bell state $\frac{\ket{00}+\ket{11}}{\sqrt{2}}$.

In summary, considering factors such as node distances, channel types, and environmental conditions, we calculate the transmissivity, as captured by \eqref{eq:fiber_t} and \eqref{eq:fso_t}. This transmissivity has a direct impact on the density matrix, as illustrated in \eqref{eq:damping}, which affects the quality of entanglement as captured by the entanglement fidelity in \eqref{eq:fidelity}. The entanglement fidelity will be used to evaluate our proposed routing strategy. Leveraging entanglement fidelity as a metric to evaluate the performance of a routing algorithm provides a quantitative measure of the entanglement quality established between the source and destination. 

\subsection{Mobility Model}
\label{sec:mobility}
\begin{figure*}
\centering
\subfloat[]{\label{fig:net_topologies_1}\includegraphics[width= 2in]{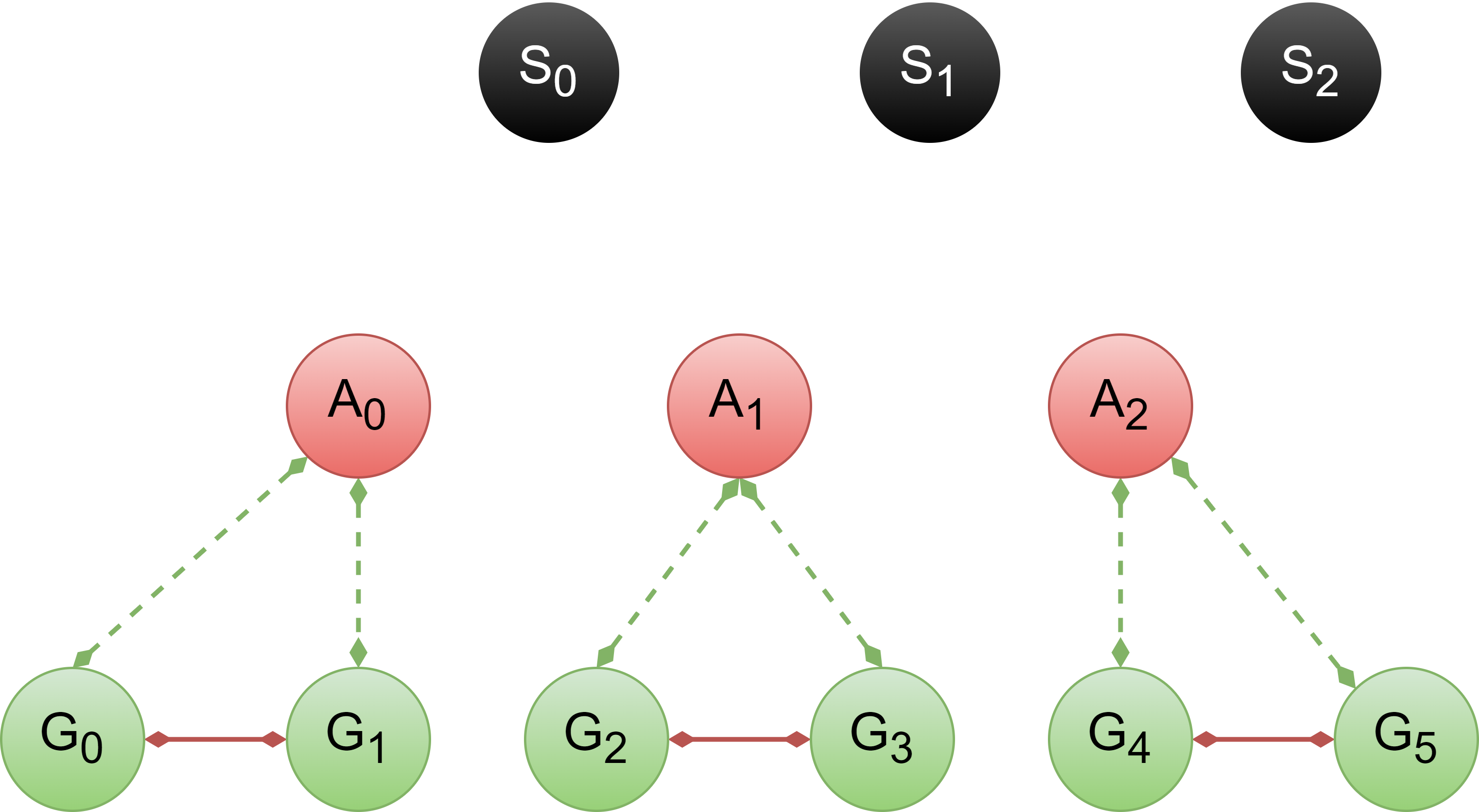}}\;\;\;\;\;\;
\subfloat[]{\label{fig:net_topologies_2}\includegraphics[width= 2in]{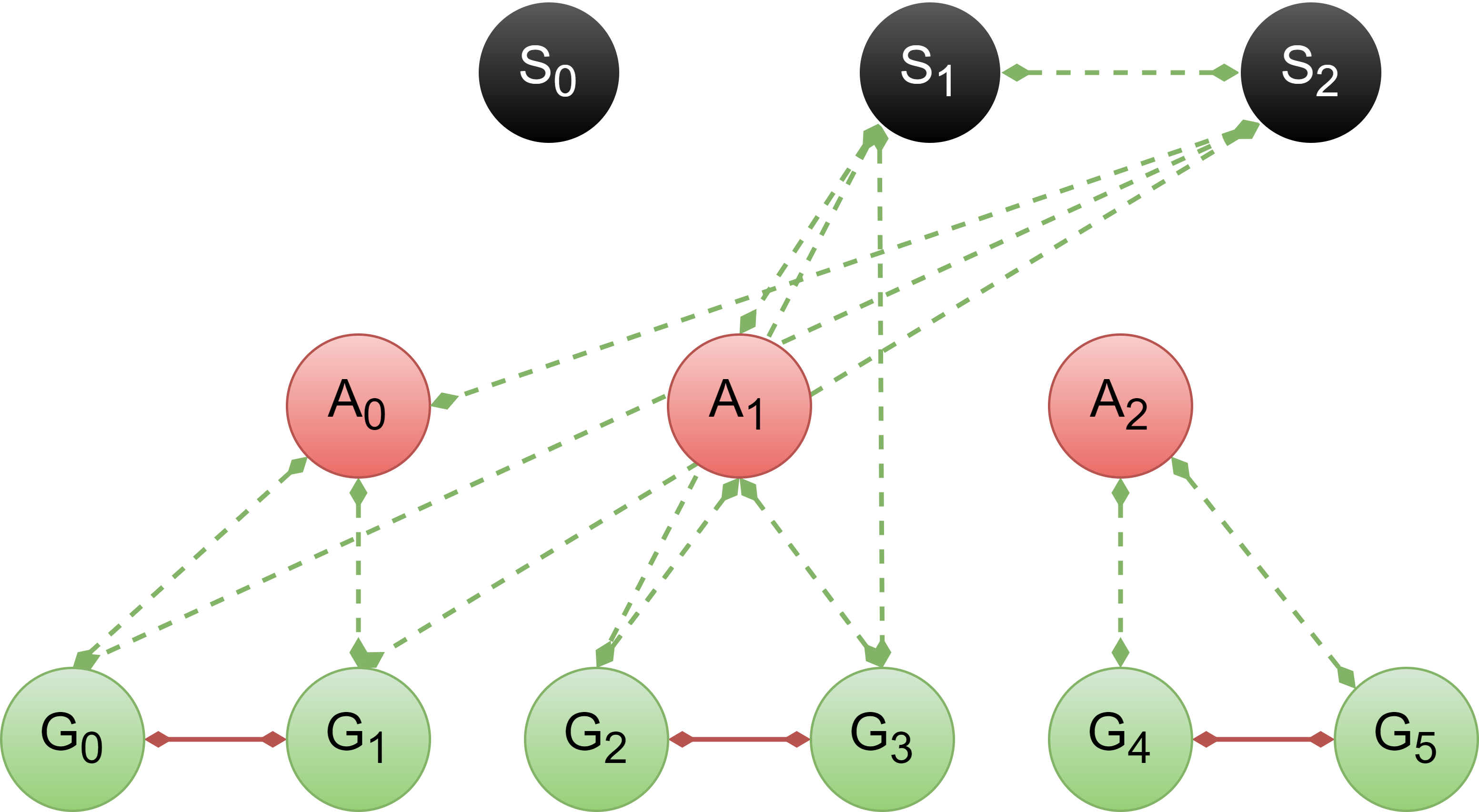}} \;\;\;\;\;\;
\subfloat[]{\label{fig:net_topologies_3}\includegraphics[width= 2in]{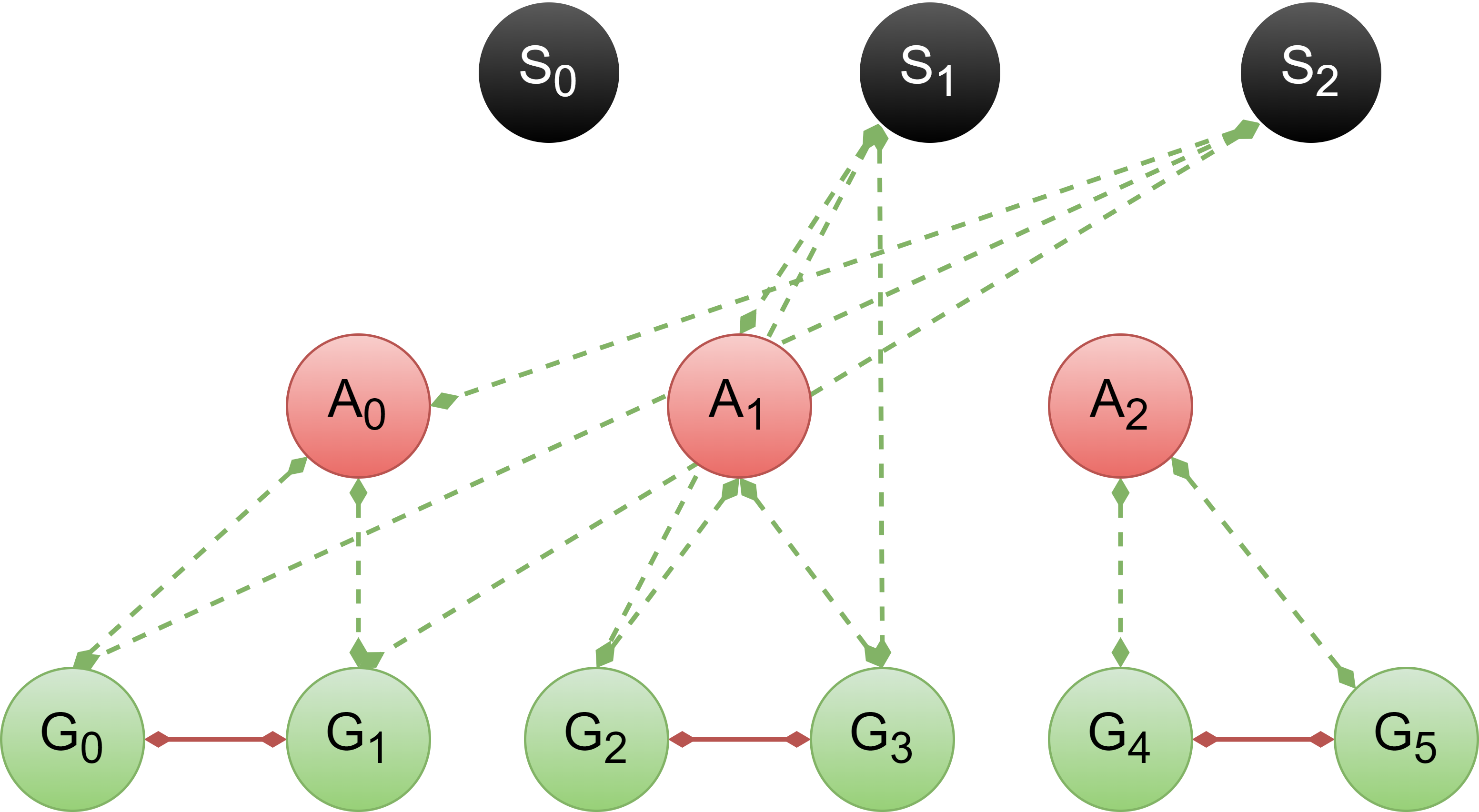}}\\
\subfloat[]{\label{fig:net_topologies_4}\includegraphics[width= 2in]{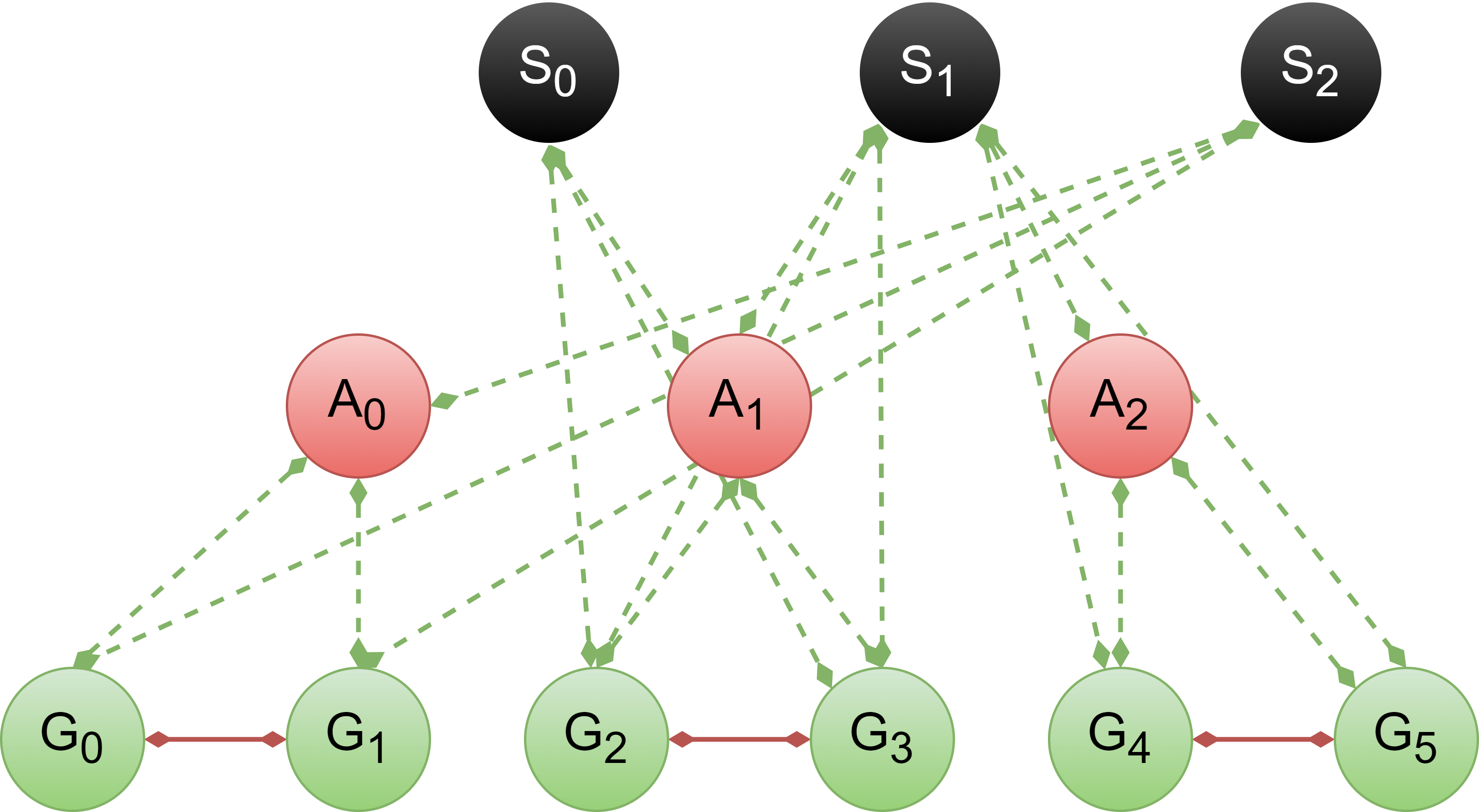}} \;\;\;\;\;\;
\subfloat[]{\label{fig:net_topologies_5}\includegraphics[width= 2in]{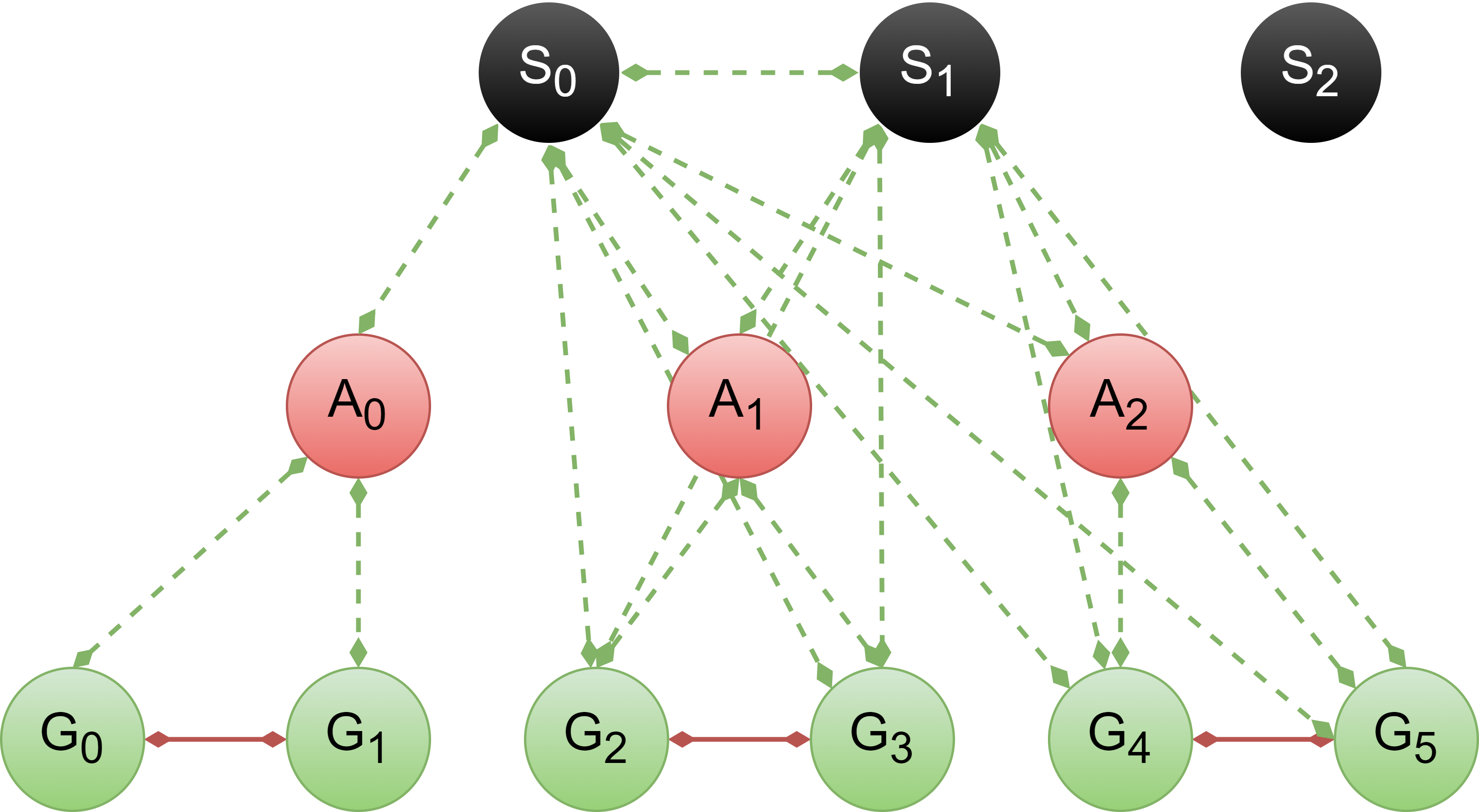}} \;\;\;\;\;\;
\subfloat[]{\label{fig:net_topologies_6}\includegraphics[width= 2in]{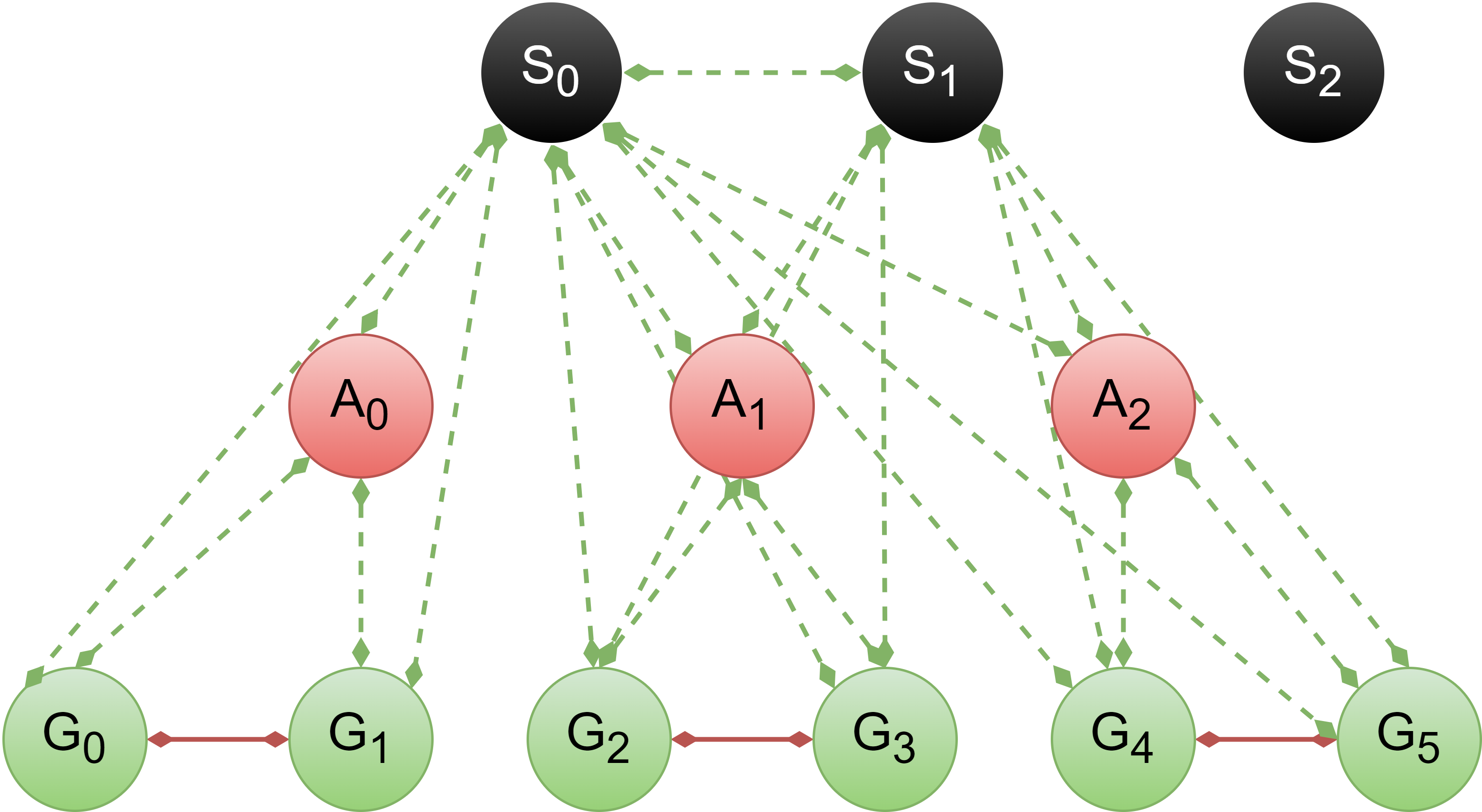}}
\caption{Illustration of different topologies of the same SPARQ network while satellites are moving. Ground nodes are presented in green, satellite nodes are presented in black, and HAP nodes are presented in red. FSO channels are presented in green dashed lines and optical fiber channels are presented in red solid lines. The shown node positions do not reflect the actual position, instead, this is reflected by the edges and whether they are connected or disconnected.}
\label{fig:net_topologies}
\end{figure*}

The mobility of satellites results in continuous changes in transmissivities, leading to dynamic connectivity and, consequently, a dynamic network topology. The STK \cite{132} simulator is employed to model the mobility of satellites. Specifically, the utilized satellites operate in LEO at an altitude of $500$ km, similar to the Micius satellite \cite{102}. It is worth noting that the STK simulator considers the Earth's motion in simulating satellite movement, thereby providing precise satellite locations. The mobility of satellites has a significant impact on the SPARQ network's topology. For instance, Fig. \ref{fig:net_topologies} provides a visual representation of various SPARQ network topologies, illustrating how the network topology changes as satellites move. Fig. \ref{fig:net_topologies} is created using the SPARQ simulator, to be detailed in Section \ref{sec:simulator}, which uses realistic satellite movements, hence, all of the parameters of this setup are practical. As shown in Fig. \ref{fig:net_topologies}, the mobility of satellites has a significant impact on the SPARQ network’s topology. This is because of the dynamic changes in transmissivities during satellite movement. In some cases, transmissivities may become too low, leading to node disconnections, while in other scenarios, transmissivity can become high and nodes get connected.  For example, in Fig. \ref{fig:net_topologies_1}, all satellites are disconnected both from each other and all network nodes due to low transmissivities. As satellites move, connectivity undergoes continuous changes based on varying transmissivities. Subsequently, in Fig. \ref{fig:net_topologies_2}, $S_1$ establishes connections with $S_2$, $A_1$, $G_2$, and $G_3$ because transmissivities between $S_1$ and these nodes become sufficiently high. However, in Fig. \ref{fig:net_topologies_3}, $S_1$ is disconnected from $S_2$, while maintaining connections with $A_1$, $G_2$, and $G_3$ due to the variable transmissivities. These connections and disconnections are determined by a transmissivity threshold of $0.7$ to support quantum teleportation and superdense coding, as will be discussed in Section \ref{sec:transmissivity_threshold}. 
 
\section{Quantum Routing within SPARQ}
\label{sec:quantum_routing}

\subsection{Problem Definition}
\label{sec:problem_definition}
We represent the SPARQ network as a graph $\mathcal{G}=(\mathcal{V},\mathcal{E})$, where $\mathcal{V}=\{v_i^x\}_{i=1}^V$ represents a set of $V$ nodes where $x \in \{\text{s},\text{a},\text{g}\}$ denoting a space, air, and ground node, respectively. Herein, $\mathcal{E}=\{e_{i,j};~v^x_i,v^x_j \in \mathcal{V}\}$ represents the edges connecting these nodes. Each edge $e_{i,j}$ can take the form of an FSO or fiber channel based on the types of nodes $v^x_i$ and $v^x_j$. Suppose node $v^\text{g}_i$ wants to transmit qubits to node $v^\text{g}_j$, and there is no direct quantum channel between these two nodes, or the direct channel yields poor transmissivity. This implies that establishing an entangled pair directly between $v^\text{g}_i$ and $v^\text{g}_j$ is not possible. In this scenario, an intermediate node $v^x_b$ that shares entangled pairs with both $v^\text{g}_i$ and $v^\text{g}_j$, meaning that $\{e_{i,b}, e_{b,j}\} \in \mathcal{E}$, can act as a repeater performing entanglement swapping to directly establish entanglement between $v^\text{g}_i$ and $v^\text{g}_j$. Therefore, the objective is to find the optimal routing path, such that, via a sequence of intermediate repeaters, a direct entanglement between communication nodes can be established with the highest possible quality. Consider the communication scenario involving the two end users represented by the black circles in Fig. \ref{fig:net}. In this case, there are several path options, and here we highlight two potential routes in yellow and black that indirectly connect these users. Assume that entangled pairs are already distributed between adjacent nodes. For the yellow and black route, each intermediate node can act as a repeater performing entanglement swapping to directly entangle the two end users. Thus, solving the routing problem involves choosing the optimal path between these users based on a specific metric for optimality. In SPARQ, the optimal path is defined as the sequence of intermediate nodes that connects the source and destination with the highest achievable entanglement quality. This is a binary optimization problem, that is NP-hard. The dynamic network topology of SPARQ, as discussed in Section \ref{sec:mobility}, necessitates solving the routing problem in real-time. This is infeasible because it is categorized as an NP-hard problem. Thus an efficient solution is needed.

\subsection{Deep RL-based Routing Strategy}
Deep RL is employed to determine the optimal route between communication nodes in SPARQ. The deep RL agent is trained using DQN. 
The use of deep RL and DQN for quantum routing within the SPARQ network is motivated by the following factors. Deep RL can simplify the solution of the routing problem, which is NP-hard as discussed in Section \ref{sec:problem_definition}, by learning through experience. In contrast, alternative approaches like stochastic optimization are impractical in SPARQ due to involving intricate computations which introduces significant delays. The requirement of computing the optimal path for each communication request further complicates the situation. Additionally, stochastic optimization approaches may converge to local minima or maxima rather than the global optimum. Furthermore, SPARQ exhibits dynamic network topology and several topologies must be captured during the training of the agent. DQN provides the adaptability required to optimize routing decisions in response to the dynamic network topology of SPARQ, ensuring efficient entanglement distribution and high-quality paths. A pivotal component of the DQN framework is the experience buffer, strategically employed to enable the agent to retain both past and recent experiences. This extensive repository of experiences equips the agent with a comprehensive understanding of the network dynamics. This comprehensive knowledge base and adaptability enable the routing decisions to be optimized in real time. The objective of the proposed DQN agent is to identify the most efficient path for entanglement distribution between nodes, considering the transmissivity between these nodes. We experimented with different hyperparameters to find the optimal settings that yield the best-performing model. We implemented the proposed DQN algorithm using a neural network with four fully connected layers. The input to the network consists of the state space $\mathcal{S}$ of the RL environment, and the output is the Q-values for each possible action. The architecture of the proposed neural network is as follows:
\begin{itemize}
    \item \emph{Input Layer:} The input layer accepts a state representation vector, which is passed through a fully connected layer. The rectified linear unit (ReLU) activation function is used to introduce non-linearity.
    \item \emph{Hidden Layers:} Two hidden layers are implemented with $128$ neurons and $64$ neurons, respectively. Also, the ReLU activation function is employed.
    \item \emph{Output Layer:} 
    The output layer consists of neurons equal in number to the possible actions in the environment, and its role is to produce the Q-values associated with each action.
\end{itemize}

\subsection{Environment and State Representation}
The environment is represented by the dynamic network graph of SPARQ. This network graph captures the current state of the network at each episode. We represent the state as a vector of transmissivities to adjacent nodes. Each node in the network is characterized by its transmissivity vector. The state vector is designed to have a fixed length, set to at least accommodate the maximum number of possible neighbors at any given node. For the cases in which a node has fewer connections, the state vector is padded with zeros to keep the fixed length. This vector serves as the state representation and is the input to the DQN agent.


\subsection{Action Space and Reward Function}
In the proposed model, the action space covers the collection of adjacent nodes to the current node. This signifies that at any given state $s_t$, the agent selects an action $a_t$ to visit one of the neighboring nodes. However, if the agent is at a ground node, its choice is constrained to prevent the next node from also being a ground node. This restriction ensures that two consecutive ground nodes do not appear in the routing path. This is essential due to the limited capabilities of ground nodes, which are unable to generate entangled states. To determine the quality of the chosen action, we use a reward function based on the transmissivity associated with the selected action. This encourages the DQN agent to favor actions that lead to higher transmissivity, thus optimizing the routing path for an efficient entanglement distribution process. The reward function is described as:
\begin{equation}
    \label{eq:reward}
        r(s,a)=
        \begin{cases}
            \text{\footnotesize{min}}[\eta_{\mbox{\tiny\itshape i,j}} \forall i,j \in \mathcal{P}] & \text{\footnotesize{if}} \; \text{\footnotesize{agent reach destination}}\\
            -1 & \text{\footnotesize{if}} \; \eta \leq 0.1 \\
            -\frac{1}{10 \times \eta} & \text{\footnotesize{otherwise}}
        \end{cases},
\end{equation} 
where $\mathcal{P}$ is the path between the source and destination nodes, which is an ordered list starting at the source and ending at the destination. As shown in \eqref{eq:reward}, the agent receives a positive reward once reaching the destination. Otherwise, for each step taken, it incurs a penalty that is inversely proportional to the transmissivity of the chosen link. This means that the agent receives a smaller penalty when choosing to pass through a channel with high transmissivity, and vice versa. Additionally, the penalty is scaled and normalized within the range of $0.1$ to $1$. The agent incurs the minimum penalty of $0.1$ when selecting a link with a transmissivity of $1$, and it incurs the maximum penalty of $1$ when selecting a link with a transmissivity of $0.1$ or lower. The second condition of \eqref{eq:reward} ensures that the penalty does not exceed $1$. The reward function in \eqref{eq:reward} is designed to prevent the agent from encountering cycles. Additionally, it encourages the agent to find an optimal path with the least number of hops. A transmissivity threshold is calculated based on acceptable entanglement fidelity value according to the application as will be shown in Section \ref{sec:transmissivity_threshold}.

\subsection{Training the DQN Agent}
Q-learning is employed to train the RL agent. The agent learns from experience by selecting an action $a$ from the action space $\mathcal{A}$, this enables the agent to acquire knowledge about the expected cumulative discounted reward $Q(s_t,a_t)$, which is known as the Q-value. In a given current state $s_t$, the agents select an action $a_t$ that offers the highest Q-value from the entire set of feasible actions within its action space, such that:
\begin{equation}
    \label{eq:lr}
    a_t=\text{arg} \max_{a_t \in A} Q(s_{t+1},a_t).
\end{equation}
We adopted the Q-learning algorithm with experience replay \cite{139}. During training, an experience replay buffer was used to store and sample experiences (state, action, reward, next state) to break the temporal correlation in the data. The loss function used for updating the Q-network was the mean squared error loss between the predicted Q-values and the target Q-values. We applied a target network, which was a duplicate of the Q-network, to stabilize the learning process. The target network parameters were updated periodically to match those of the Q-network. The DQN was trained using mini-batch stochastic gradient descent. The Q-values are updated as follows:  
\begin{align}
 \nonumber Q(s_t,a_t)&= ( 1-\beta) Q(s_t,a_t) \\
 &+\beta \left(r(s_t,a_t)+\gamma \max_{a_t \in \mathcal{A}} Q(s_{t+1},a_t) \right),
\end{align}
where $\beta$ is the learning rate, $\gamma$ is the discount factor, $s_t$ is the current state, $a_t$ is the action chosen by the agent, and $r$ is the reward function. We experimented with various hyperparameters, including the discount factor $(\gamma)$, exploration rate $(\epsilon)$, batch size, and replay buffer size, to find the optimal settings for our specific environment \textcolor{black}{(see Appendix \ref{app:sim})}.

\begin{figure*}
    \centering
    \includegraphics[width=7in]{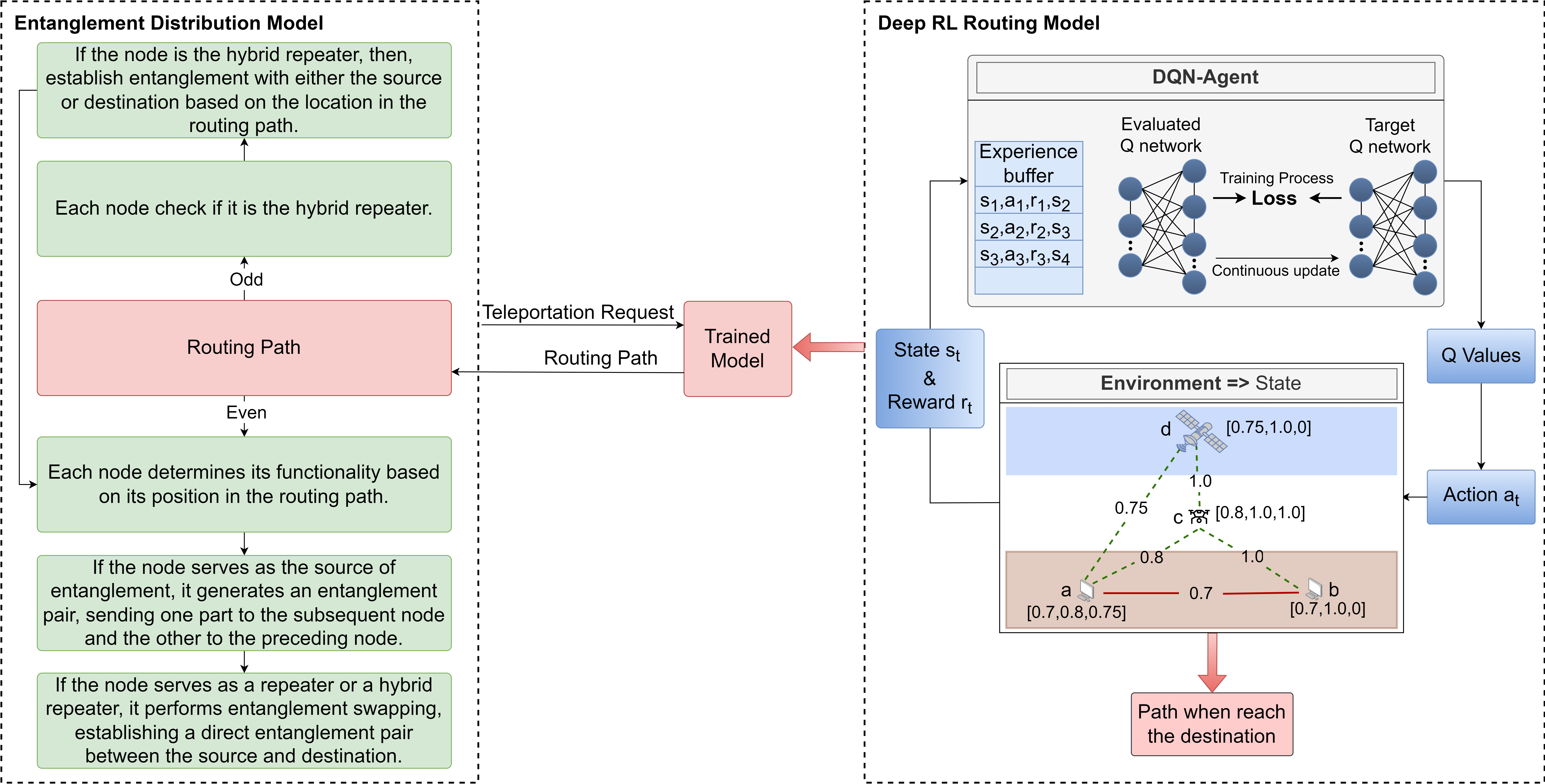}
    \caption{Illustration of the proposed routing model and entanglement distribution strategy.
    The TPED policy is employed here for entanglement distribution as it results in improved end-to-end entanglement quality, as will be detailed in Section \ref{sec:simulation_results}.}
    \label{fig:block_diagram}
\end{figure*}

The SPARQ network graph is dynamic as it experiences changes in the positions and connectivity of satellites over time. At each training episode, we update the environment to capture the current state of the network graph. This ensures that the DQN agent adapts to the dynamic network topology of SPARQ. Simulation results indicate the superiority of this approach compared to training the agent on a single snapshot of SPARQ, as detailed in Section \ref{sec:results}. Our DQN agent undergoes $500$ training episodes, and the following steps are performed for each episode:
    \begin{enumerate}
        \item The environment (network graph) is updated to reflect the current state of the SPARQ network.
        \item Edge features, represented by transmissivities, are transformed into a vector at each node, forming node features. This step allows us to represent the state space and incorporate it into our DQN model.
        \item The DQN agent is trained for $100$ additional episodes (mini-episodes) on the current graph.
    \end{enumerate}
This methodology outlines the approach used to design and train the DQN agent for routing optimization within the dynamic SPARQ network. It provides details of how the model interacts with the environment, how state representations are created, and how the model's policy is developed to determine optimal routing paths based on transmissivities. The proposed routing strategy is illustrated on the right side of Fig. \ref{fig:block_diagram}. We employ offline training to train our agent, and once the training is complete, we deploy the agent at each node in the network. At each node, the agent relies solely on local information to determine the next node for entanglement distribution. Therefore, there is no need for centralized coordination. Specifically, the agent uses the transmissivities to adjacent nodes to determine the encountered network state and select the next node that maximizes the cumulative reward. Given that satellite trajectories are periodic, this local information effectively reflects the network state, enabling the agent to optimize routing requests based on encountered network states. Once the destination is reached, the selected routing path is shared with each node involved to follow an entanglement distribution policy. This decentralized approach eliminates the need for a central controller. Each node is equipped with its own trained agent to make real-time decisions based on local information, ensuring scalability and feasibility in a dynamic SPARQ environment. \textcolor{black}{Further details on the dataset employed for training the DQN agent can be found in Appendix \ref{app:dataset}.}

\subsection{Convergence}
The proposed DQN algorithm's convergence is tracked by observing the loss function during the training process. The training is halted when the difference in loss between epochs reaches a sufficiently low value ($5\times10^{-4}$) for $50$ consecutive epochs. Observing minimal differences between successive epochs indicates stability in the learning process. When the algorithm converges, the Q-network's updates become progressively marginal, signifying that the model has learned a stable policy.

\subsection{Complexity}
The time complexity of a deep RL algorithm with a state space $\mathcal{S}$, action space $\mathcal{A}$, exploration rate $\epsilon$, and discount factor $\gamma$ can be expressed as $O(|\mathcal{S}|^2|\mathcal{A}|/(\epsilon^3(1-\gamma)^3))$ \cite{133}, where $|.|$ denotes the cardinality of a space. Here, considering a SPARQ network with $V$ nodes, the time complexity of the proposed deeper RL algorithm is upper bounded by $O(V^3/(\epsilon^3(1-\gamma)^3))$, denoting polynomial complexity, which is considered efficient.

\section{Entanglement Distribution within SPARQ}
\label{sec:entanglement_distribution}
\begin{figure*}
\centering
\subfloat[]{\includegraphics[width= 2in]{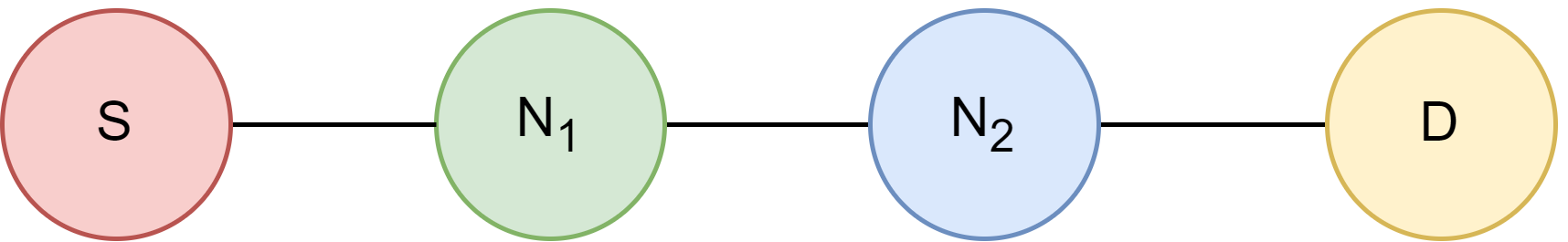}} \;\;\;\;\;\;
\subfloat[]{\includegraphics[width= 2in]{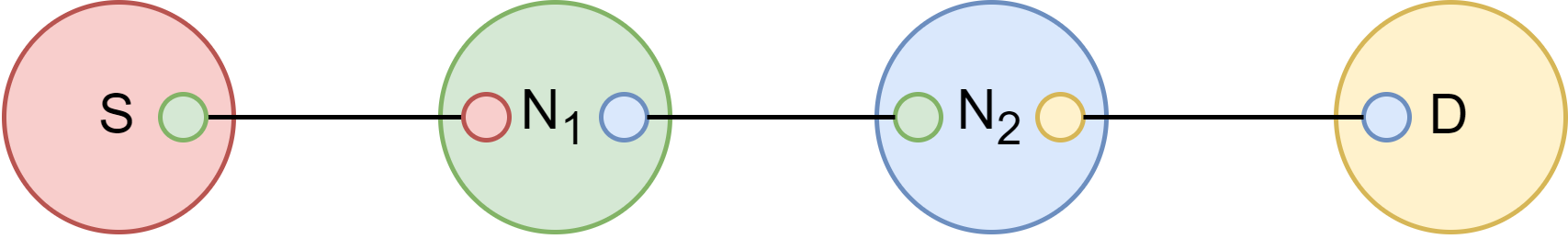}\label{ED_FP}} \;\;\;\;\;\;
\subfloat[]{\includegraphics[width= 2in]{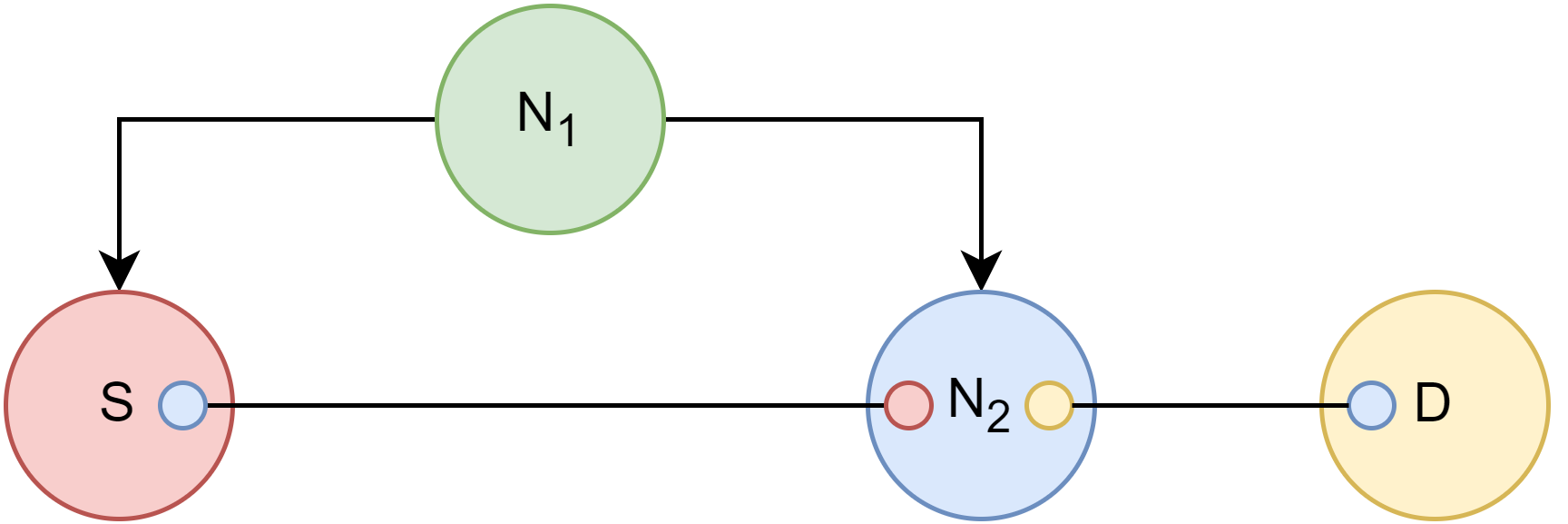}\label{ED_TP}}
\caption{Two different policies to establish entanglement between the source node $S$ and the destination node $D$, the circles inside nodes represent an entanglement state. The color of the entanglement state represents the node with which an entanglement state is shared. In this figure: (a) represents the network nodes and communication channels before establishing entanglement, (b) represents the intuitive approach, and (c) represents the proposed TPED policy, where the role of $N_1$ is to generate an entangled pair and send one part to $S$ and the other part to $N_2$.}
\label{fig:ED_policies}
\end{figure*}

Once the routing path between the source and destination is determined, the subsequent phase involves establishing entanglement between them, utilizing intermediate nodes according to a predefined policy. As mentioned earlier, existing entanglement distribution policies are not applicable in SPARQ as they assume uniform functionalities across network nodes. In SPARQ, not all network nodes possess the capability to generate entanglement pairs. The intuitive approach for entanglement distribution is to establish entanglement between adjacent nodes, considering network constraints such that satellites and HAPs can serve as sources of entanglement, while ground nodes lack the ability to generate entangled pairs. Intermediate nodes then execute entanglement swapping operations to establish direct entanglement between the communication parties. However, this approach leads to a degradation in the end-to-end entanglement shared between the source and destination. This is because each entanglement swapping process degrades the entanglement quality due to non-maximally entangled states being shared between adjacent nodes \cite{105}. Therefore, we propose an entanglement distribution policy, namely TPED. The TPED policy aims to minimize the number of entangled pairs consumed by the intermediate repeaters. This policy seeks to enhance the overall entanglement quality of the end-to-end entangled pair established between the source and destination. The idea of the proposed TPED policy is based on keeping the source of entanglement outside the routing path. Each source of entanglement can achieve this by generating an entangled pair and sending one part to the preceding node in the routing path, while the other part is sent to the subsequent node. This allows for the adjacent nodes of the source of entanglement to get entangled together directly. The difference between the intuitive approach and the TPED policy is shown in Fig. \ref{fig:ED_policies}. In this figure, the entanglement distribution process can be performed in different ways. For instance, $N_1$ can establish entanglement with $S$ and $N_2$, then, $N_2$ can establish entanglement with $D$. Now, we have a chain of entanglement ($S 	\longleftrightarrow N_1 	\longleftrightarrow N_2 	\longleftrightarrow D$) between $S$ and $D$ as shown in Fig. \ref{ED_FP}. By performing entanglement swapping at nodes $N_1$ and $N_2$, $S$ and $D$ share a direct entangled pair that can be used for communication. This approach requires creating three entangled pairs and performing two swapping operations to establish entanglement between $S$ and $D$. In TPED, we consider generating an entangled pair at $N_1$, then sending half to $S$ and the other half to $N_2$. Meanwhile, $N_2$ establishes an entanglement connection with $D$. Now, we have a chain of entanglement ($S 	\longleftrightarrow N_2 	\longleftrightarrow D$) between $S$ and $D$ shown in Fig. \ref{ED_TP}. By performing entanglement swapping at nodes $N_2$, $S$ and $D$ share a direct entangled pair that can be used for communication. This approach only requires two entangled pairs and a single entanglement swapping operation to establish entanglement between $S$ and $D$. 

The TPED assigns functionalities to each node along the routing path to establish entanglement between communication parties while minimizing the number of entanglement swaps. This approach also considers node limitations, such as the inability of ground nodes to serve as sources of entanglement. The TPED policy depends on the number of nodes involved in the communication. If the number of nodes in the routing path is odd, the policy will follow the following steps:
\begin{enumerate}
    \item Start from the first intermediate node, which is the node after the source, to the last intermediate node, which is the node before the destination. In an alternating pattern, each node is marked as either a source of entanglement or a repeater, respectively. 
    \item Each source of entanglement generates an entangled pair and sends half to the preceding node and the other half to the subsequent node of the routing path. Hence, the preceding and the subsequent nodes are connected by this source of entanglement. 
    \item Each repeater performs entanglement swapping to directly entangle the source and the destination.
\end{enumerate}

If the number of hops is even, the policy designates one intermediate node to serve a dual role, acting both as a source of entanglement and as a repeater. This node is referred to as the hybrid repeater. Meanwhile, the remaining intermediate nodes continue to serve as either sources of entanglement or repeaters. It is worth mentioning that the SPARQ network limits the functionality of the ground nodes to either end-node or repeaters. Therefore, ground nodes cannot be used as a source of entanglement. Then, the policy goes as follows:
\begin{enumerate}
    \item The initial step is to select the hybrid repeater, where the role of the hybrid repeater is assigned only to either the first or the last intermediate node of the routing path. The selection is based on the type of the two nodes that they connect in the routing path. One of these nodes is a ground node (source or destination) and the other can be ground, satellite, or HAP. The intermediate node is selected as the hybrid repeater if it connects one ground node (source or destination) with one source of entanglement (satellite or HAP) in the routing path. In cases where both the first and last intermediate nodes are eligible for selection as the hybrid repeater, either of them can be chosen. These restrictions on the choice of the hybrid repeater allow the network nodes to follow a deterministic policy during entanglement distribution.
    \item Start from the first intermediate node to the last intermediate in the routing path. In an alternating pattern, each node except for the hybrid repeater is marked as either a source of entanglement or a repeater, respectively. 
    \item Each source of entanglement generates an entangled pair and sends half to the preceding node and the other half to the subsequent node of the routing path. Hence, the preceding and the subsequent nodes are connected by this source of entanglement. Also, the hybrid repeater establishes entanglement with either the source or the destination based on the selection of the hybrid repeater.  
    \item Each repeater and the hybrid repeater perform entanglement swapping to establish a direct entanglement between the source and the destination. 
\end{enumerate}
The whole process with the proposed entanglement distribution based on the TPED policy is illustrated in Fig. \ref{fig:block_diagram}.

\section{Simulation Results and Analysis}
\label{sec:results}
\subsection{SPARQ Simulator}
\label{sec:simulator}
Currently, quantum network simulators, such as QuNetSim \cite{41}, NetSquid \cite{117}, QDNS \cite{118}, SQUANCH \cite{120}, SeQUeNCe \cite{121}, SimulaQron \cite{122}, and SimQN \cite{10024900} primarily focus on ground networks and terrestrial communications. There is no existing quantum network simulator that can be used to carry out SPARQ's simulation. To bridge this gap, we have upgraded the QuNetSim simulator to evaluate the performance of the proposed SPARQ's architecture and the proposed routing model. This upgraded version of the QuNetSim simulator is called SPARQ simulator. We chose QuNetSim because it is easy to configure and upgrade, integrates with multiple backends, and includes several built-in protocols such as teleportation and entanglement swapping. Also, it utilizes graphs to represent the network, supports the creation of separate networks for classical and quantum communications, and offers flexibility in developing and testing routing algorithms for both classical and quantum networks. \textcolor{black}{Here, we note that QuNetSim is an abstract quantum network simulator, focusing more on higher-level quantum networking protocols than on the detailed physical layer. As a result, it lacks detailed modeling of specific hardware characteristics, such as the imperfections and noise associated with quantum repeaters, detectors, or entanglement sources. Additionally, the simulator assumes perfect quantum memory, where quantum states can be stored without degradation.} Our SPARQ simulator integrates QuNetSim and STK \cite{132} to account for realistic space-air-ground node behaviors and models. 
While QuNetSim already features an implementation of fiber optic channels, it lacks the implementation of FSO channels. Hence, an FSO channel model was implemented as described in Section \ref{sec:fso}. This implementation involves creating a new class within the simulator for FSO channels. This class holds the FSO channel properties, which are configured using the constructor. Furthermore, a function is defined within the class to calculate and store the transmissivity according to \eqref{eq:fso_t}. Additionally, various functions are implemented to calculate the equations necessary for determining the transmissivity. These functions are invoked by the primary transmissivity calculation function. Notably, the setter for the distance recalculates the transmissivity for this channel. Therefore, adjusting the channel distance during satellite motion results in updating the transmissivity. This ensures the accurate computation of transmissivity for satellite links during movement. Moreover, QuNetSim defines the \textit{Host} class for representing ground nodes within quantum networks. We extended this class to include information about node location such as latitude, longitude, and altitude. Furthermore, new classes were implemented to model HAPs and satellites within the proposed SPARQ network. These classes extend the \textit{Host} class and implement specific properties relevant to each host type. For example, the \textit{Satellite} class includes a movement list detailing the satellite's successive locations. Additionally, we define a thread for updating the satellite's location. This thread adjusts the satellite's current location to the next entry in the movement list. During each step of satellite movement, the satellite updates the distance to each connected node, which results in updated transmissivity. Furthermore, new functions were defined to degrade the quality of entangled states as described in \eqref{eq:damping}, and measure the quality of entanglement using the entanglement fidelity as defined by \eqref{eq:fidelity}.  Furthermore, the STK simulator is used to simulate the movement of satellites. Each satellite is created in its orbit, ensuring a uniform distribution of orbits around the Earth. Subsequently, the simulation is executed to model the movement of satellites throughout one day, recording the location of each satellite at intervals of every $30$ seconds. These recorded locations are used to generate movement sheets which include precise locations of each satellite while moving including latitude, longitude, and altitude. These movement sheets are then employed to simulate satellite movements within the SPARQ simulator. Specifically, when creating a satellite node within SPARQ, the corresponding sheet for that satellite is converted to a movement list. As the satellite navigates within the network, its current location is continuously updated to the subsequent entry in the movement list as described earlier.

\subsection{Identifying the Transmissivity Threshold}
\label{sec:transmissivity_threshold}
\begin{figure}
    \centering
    \includegraphics[width=2.9in]{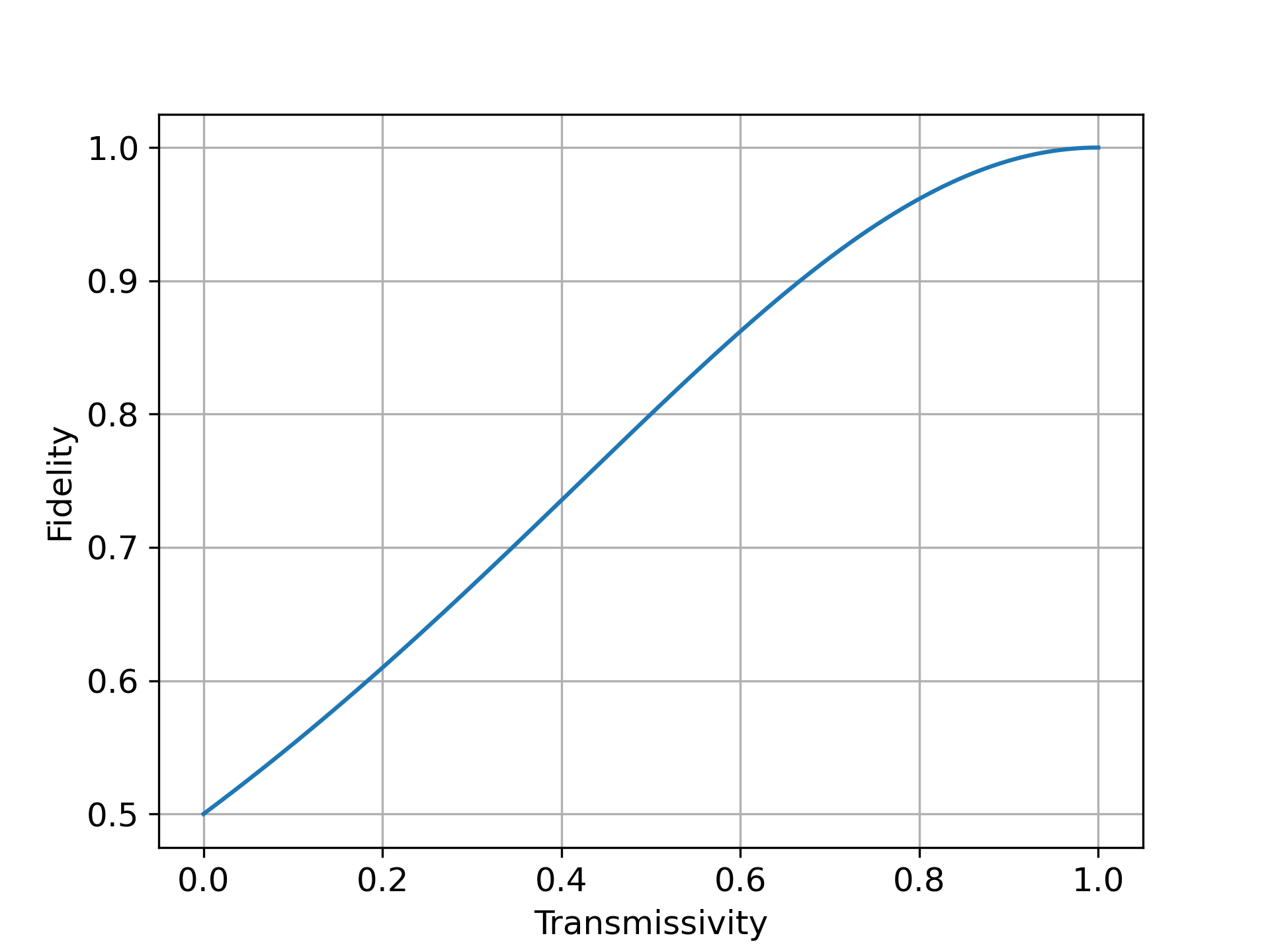}
    \caption{Relationship between transmissivity and entanglement fidelity.}
    \label{fig:EOF_TP}
\end{figure}

In Section \ref{sec:mobility}, we indicated that a transmissivity threshold is needed to decide the SPARQ topology, and hence, identify a routing path following our proposed RL-based strategy. Here, we conducted an experiment using the SPARQ simulator to determine this transmissivity threshold that can be used for quantum communication. This experiment aims to show the relationship between the transmissivity and the entanglement fidelity. To achieve this, we established different quantum links with different transmissivity ranging from $0$ to $1$ in a step of $0.01$. Regardless of the node and communication channel types, the entanglement fidelity remains consistent for the same value of transmissivity, as described in \eqref{eq:kraus} and \eqref{eq:damping}. Therefore, for simplicity, a scenario with two ground nodes connected by a fiber optical channel is considered in this experiment. To determine the channel's parameters that should be used to achieve the desired transmissivity ranging from $0$ to $1$, we solve \eqref{eq:fiber_t} setting the attenuation coefficient $\alpha$ to $1$ for simplicity. Hence, for each channel, without loss of generality, the attenuation coefficient $\alpha$ is set to $1$, and the distance between the two nodes is set to $-\ln(x)$, where $x$ is the desired transmissivity. In this manner, mobility is implicitly considered. Then, an entangled pair is established across each link and the fidelity is measured for each entanglement. The relationship is shown in Fig. \ref{fig:EOF_TP}.

As shown in Fig. \ref{fig:EOF_TP}, a transmissivity of $0.7$ results in an entanglement fidelity exceeding $90\%$, which can enable high-fidelity teleportation and hence, exchanging quantum information \cite{134,135}. Therefore, the transmissivity threshold is set to $0.7$. The transmissivity threshold is the limit that determines whether transmissivity is sufficient for enabling teleportation and, hence, allowing quantum communications. It serves as a criterion for establishing connectivity between nodes within SPARQ, determining whether they are effectively connected or disconnected. A link is established between two nodes if its transmissivity is at least equal to the specified threshold. For instance, during satellite movements, transmissivities undergo continuous changes, leading to dynamic connections and disconnections of links based on the transmissivity threshold.

\begin{figure}
    \centering
    \includegraphics[width=3.3in]{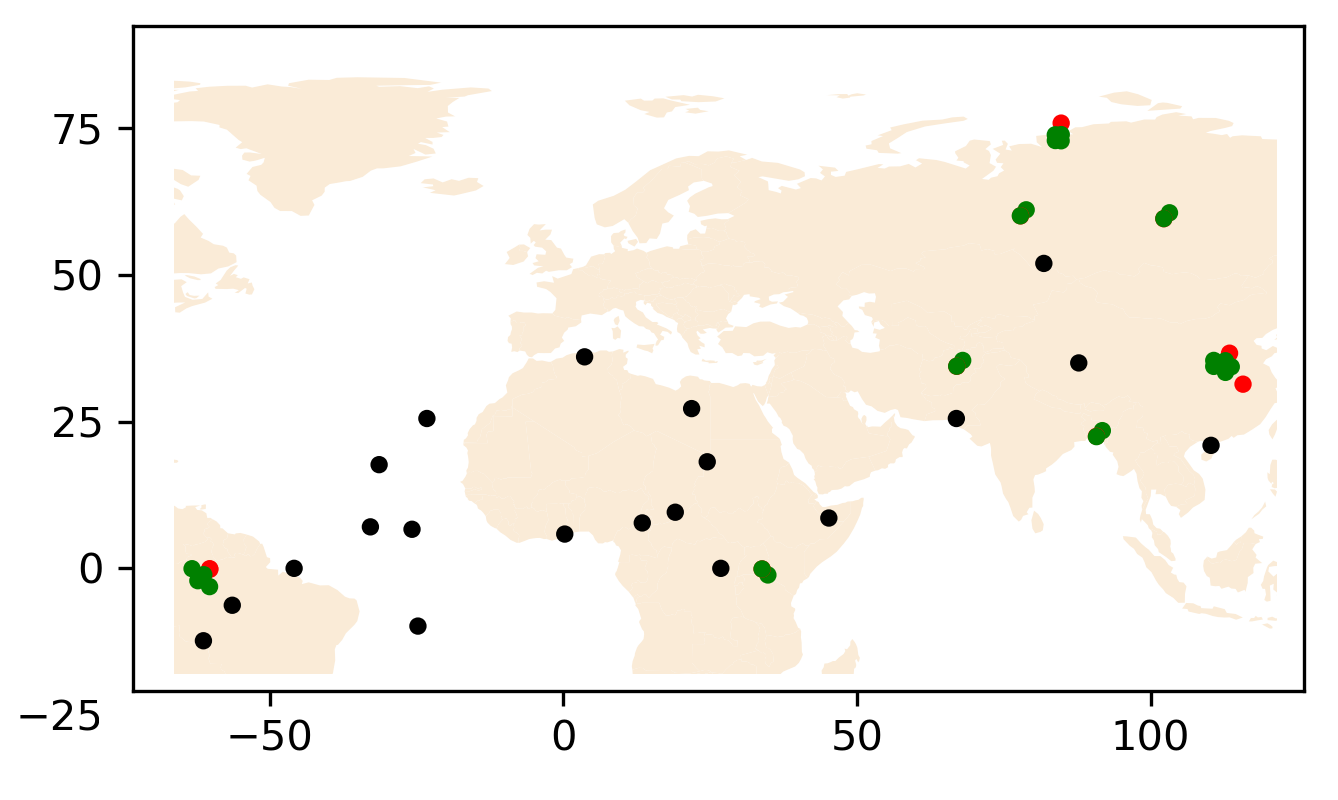}
    \caption{Setup of the SPARQ network used for simulation. Ground nodes are presented in green, satellite nodes are presented in black, and HAP nodes are presented in red.}
    \label{fig:net_setup}
\end{figure}

\subsection{Simulation Setup}
As a proof of concept, the proposed DQN-based routing strategy is tested in a SPARQ network with $54$ nodes, containing $24$ ground nodes, $20$ satellites, and $10$ HAPs. The satellites operate in LEO at an altitude of $500$ km, similar to the Micius satellite \cite{102}. This altitude is chosen to maximize the satellites' lifetime while minimizing signal loss \cite{king1987prediction}. The HAPs are positioned at an altitude of $50$ km, which is the maximum flying altitude defined by the International Telecommunication Union (ITU) for HAPs, ranging from $20$ to $50$ km \cite{9356529}. We selected the maximum altitude for HAPs to minimize the length of satellite links, thereby enhancing link quality and coverage. \textcolor{black}{The simulation parameters used in our experiments are summarized in Appendix \ref{app:sim}}. The network setup is shown in Fig. \ref{fig:net_setup} where ground nodes are presented in green, satellite nodes are presented in black, and HAP nodes are presented in red. The node locations are randomly selected by generating random latitude and longitude coordinates, ensuring through the Geopy \cite{131} library that the generated locations are not over the sea. Moreover, some nodes are manually positioned to bring them into proximity with other nodes, forming local networks. \textcolor{black}{Detailed information regarding the coordinates of these nodes and the configuration of satellites is provided in Appendix \ref{app:config}.} The network architecture is designed to have several local networks separated by large distances across several countries. Each local network contains a few ground nodes connected using fiber optical channels. Communication links are established between ground nodes if the transmissivity is at least $0.7$. HAPs may exist to extend the communication distance between network nodes and fulfill teleportation requests whenever satellites are not in range. Similarly, communication links are established among HAPs and between HAPs and ground nodes if the transmissivity is at least $0.7$. Each satellite follows its orbital trajectory, dynamically connecting and disconnecting from network nodes based on the transmissivity threshold. Notably, this architecture serves both classical and quantum data transmission, as quantum channels can be used for transmitting both types of data \cite{5592851}. We chose this architecture to simulate the early-stage scenario of quantum networks where the quantum nodes will be distributed across vast geographical areas. For instance, IBM is working to establish quantum computers across the globe, including locations such as Germany, Japan, South Korea, and Canada in addition to existing quantum computers in the United States \cite{123,124,125}.

\begin{figure}
    \centering
    \includegraphics[width=2.8in]{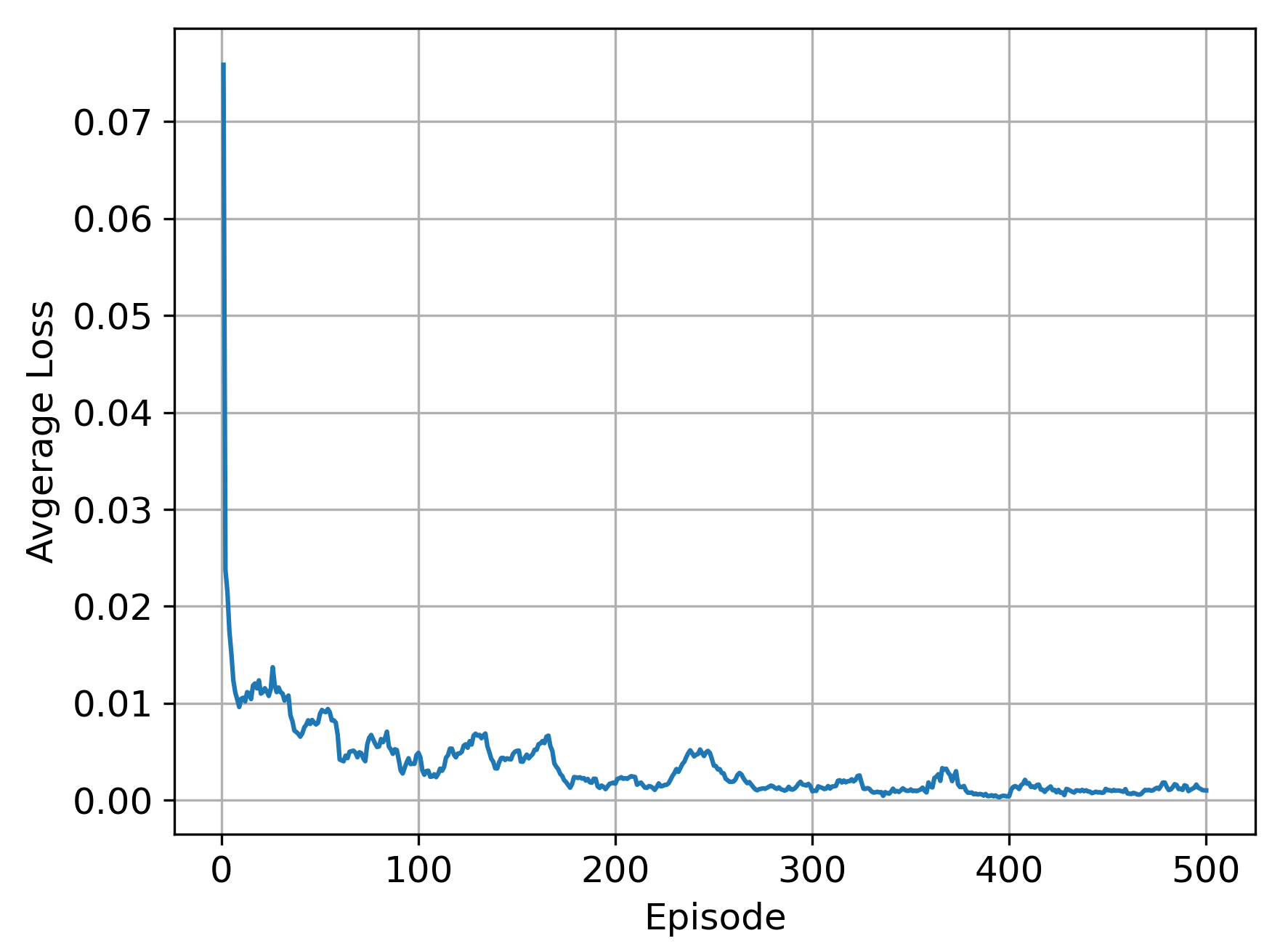}
    \caption{Average training loss of the proposed DQN algorithm.}
    \label{fig:loss}
\end{figure}

\subsection{Simulation Results}
\label{sec:simulation_results}
\begin{figure}
    \centering
    \captionsetup{labelfont={color=black}}
    \includegraphics[width=3in]{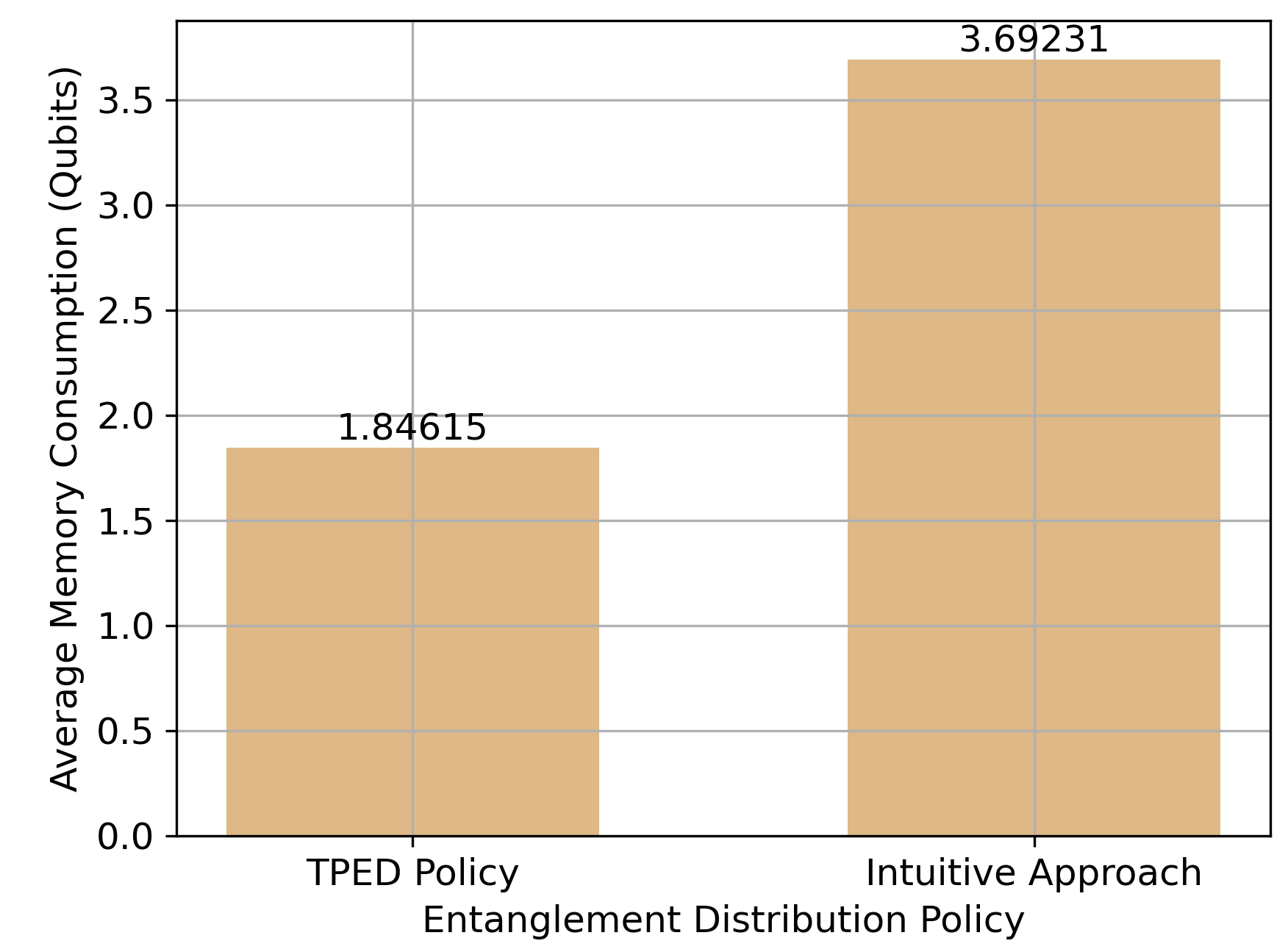}
    \caption{Comparing the TPED policy with the intuitive approach in terms of the average memory consumption.}
    \label{fig:TPED_FPED_Memory}
\end{figure}

Fig. \ref{fig:TPED_FPED} compares the performance of the proposed TPED policy for on-demand entanglement distribution with the intuitive approach. The time step reflects the movements of the satellites. In this experiment, the communication parties are chosen randomly from the ground nodes, the proposed DQN algorithm is used to find the optimal routing path, and each entanglement distribution policy is used separately to establish on-demand end-to-end entanglement between the communication parties. The entanglement fidelity is measured for each entangled pair established by each policy. The experiment is repeated $100$ times and the average is recorded for each time step over $100$ moves of satellites. As shown in Fig. \ref{fig:TPED_FPED}, the TPED policy outperforms the intuitive approach. Specifically, the TPED policy achieved an average entanglement fidelity of $0.988$, while the intuitive approach achieved an average entanglement fidelity of $0.96$. This means that the TPED policy improved the average entanglement fidelity by about $3\%$ compared with the intuitive approach. This is because TPED minimizes the number of utilized entangled pairs and minimizes the number of entanglement swapping operations performed. Furthermore, Fig. \ref{fig:TPED_FPED_Memory} compares the TPED policy and the intuitive approach in terms of average memory units consumed by intermediate repeaters. In the scenario of Fig. \ref{fig:TPED_FPED_Memory}, the source and destination nodes are randomly selected from the ground nodes. Each policy is then employed to distribute entanglement between the source and destination, with the number of memory units consumed by intermediate repeaters being recorded. The experiment is repeated $100$ times, and the average memory consumption is calculated. This improvement is attributed to the efficiency of the proposed TPED policy, which minimizes the number of utilized entangled pairs. As shown in Fig. \ref{fig:TPED_FPED_Memory}, the TPED policy reduces memory consumption by $50\%$. As such, the TPED policy will be adopted in all experiments.

{\color{black}\begin{figure}
    \centering
    \captionsetup{labelfont={color=black}}
    \includegraphics[width=3in]{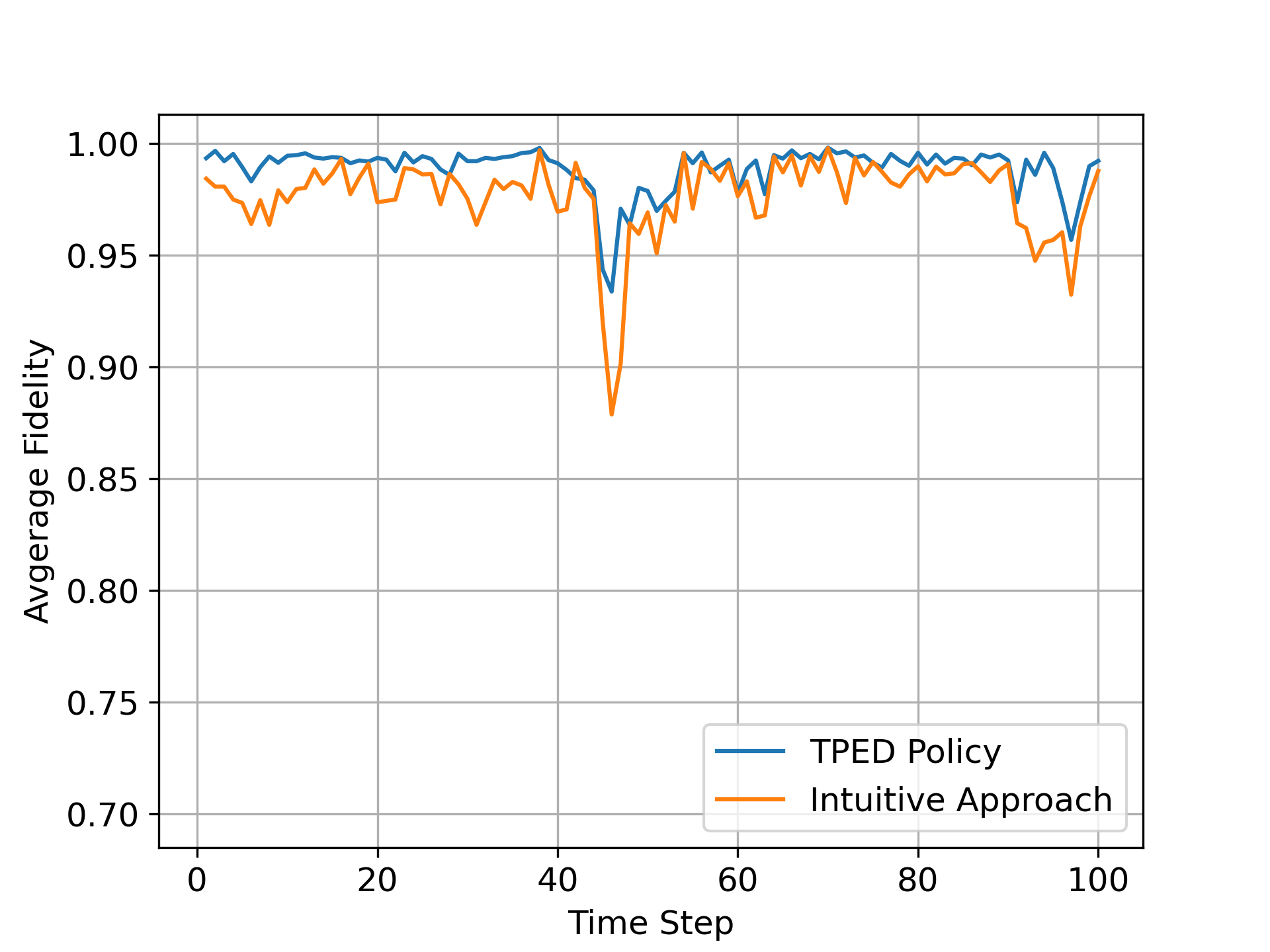}
    \caption{Comparing the TPED policy with the intuitive approach in terms of the average entanglement fidelity.}
    \label{fig:TPED_FPED}
\end{figure}}

\begin{figure}
    \centering
    \captionsetup{labelfont={color=black}}
    \includegraphics[width=3in]{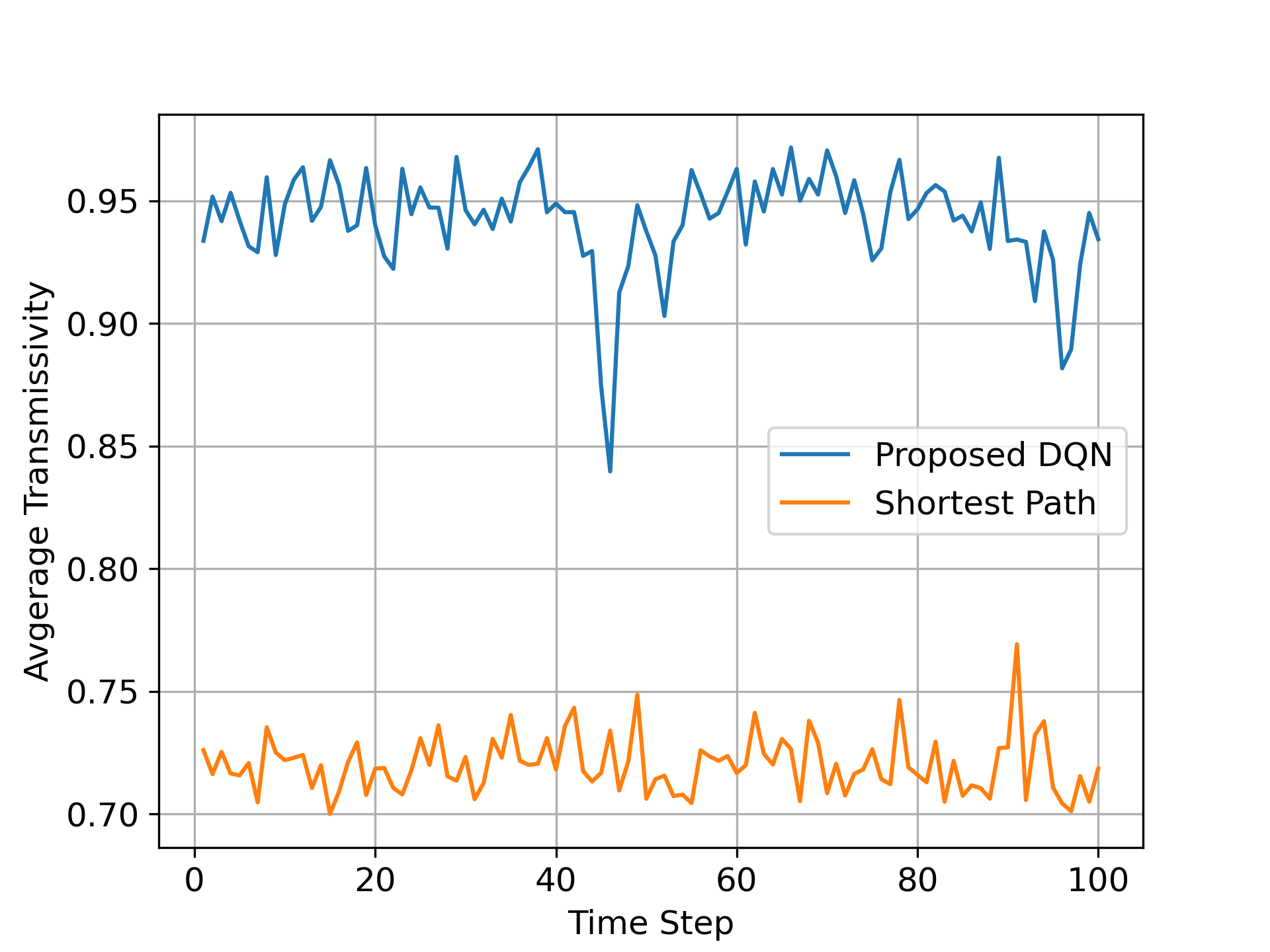}
    \caption{Comparing the proposed DQN algorithm with the shortest path algorithm in terms of the average transmissivity.}
    \label{fig:DQN_vs_Shortest_TP}
\end{figure}

\begin{figure}
    \centering
    \captionsetup{labelfont={color=black}}
    \includegraphics[width=3in]{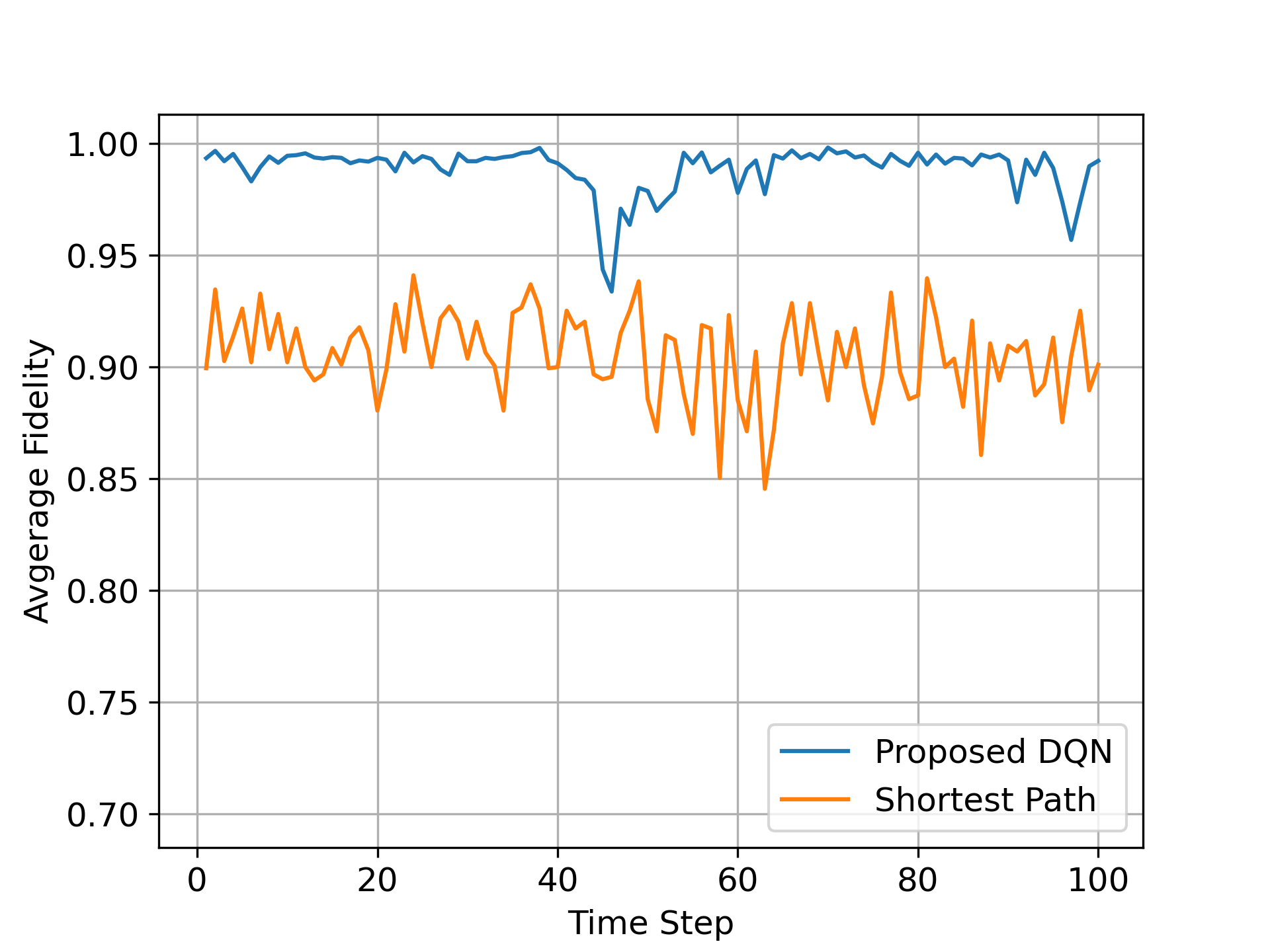}
    \caption{Comparing the proposed DQN algorithm with the shortest path algorithm in terms of the average entanglement fidelity.}
    \label{fig:DQN_vs_Shortest_EOF}
\end{figure}

\begin{figure}
    \centering
    \captionsetup{labelfont={color=black}}
    \includegraphics[width=3in]{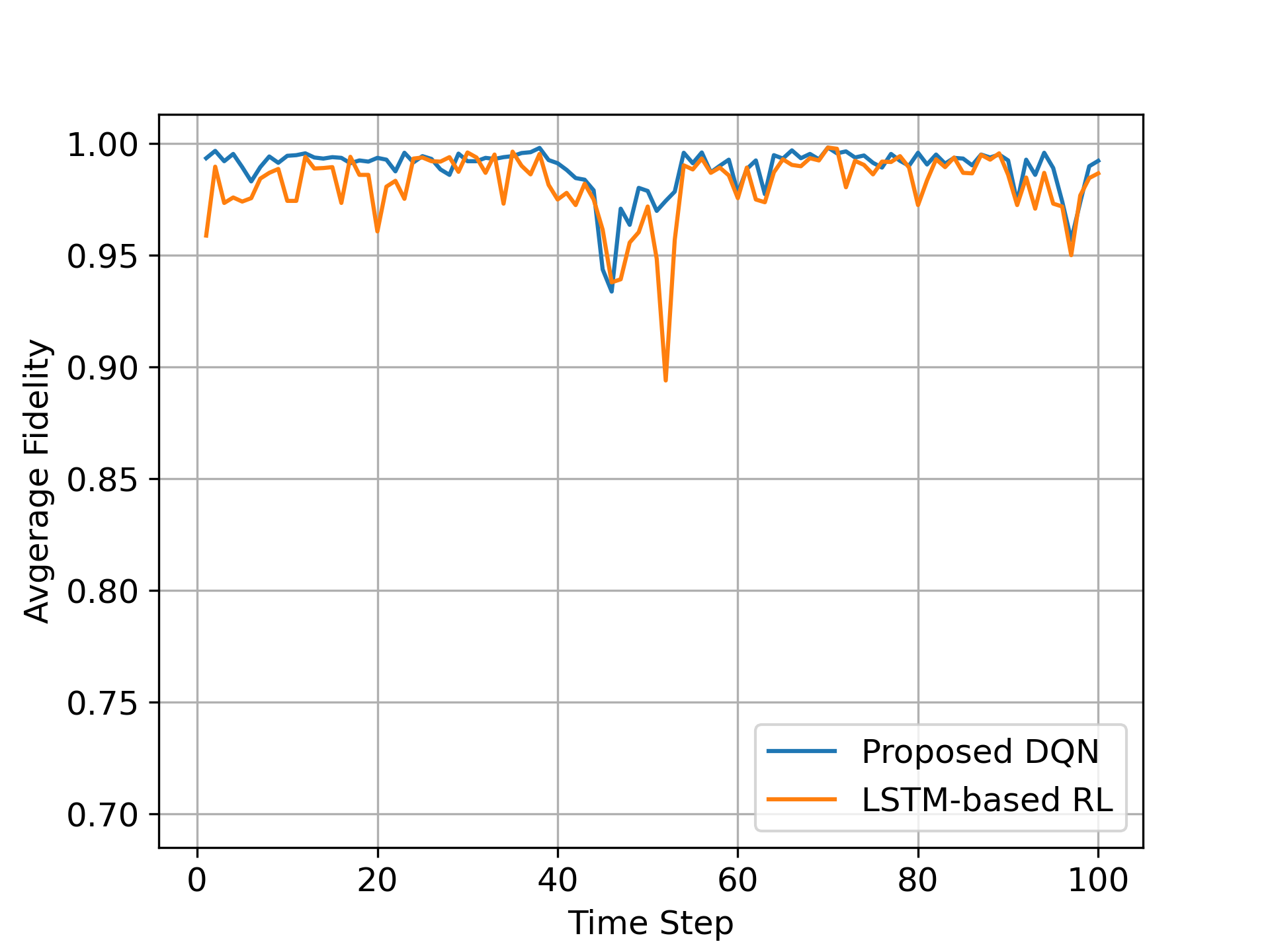}
    \caption{Comparing the proposed DQN algorithm with the LSTM-based RL algorithm in terms of the average entanglement fidelity.}
    \label{fig:DQN_vs_LSTM_EOF}
\end{figure}

Fig. \ref{fig:DQN_vs_Shortest_TP} compares the proposed DQN algorithm with the shortest path algorithm in terms of the average transmissivity while satellites are moving. In this experiment, the shortest path algorithm is selected as a baseline for performance evaluation due to its extensive use in the literature for addressing the quantum routing problem \cite{27,24,25,21,23,111,112,113,114,115}. In order to conduct this experiment, the source and destination are chosen randomly from the ground nodes. Then, the proposed DQN scheme is used to find the optimal path between the source and the destination over $100$ steps of satellites. Also, the Dijkstra algorithm is used to find the shortest path between the two parties involved in the communication. The experiment is repeated $100$ times and the average is recorded. As shown in Fig. \ref{fig:DQN_vs_Shortest_TP}, the proposed DQN algorithm outperforms the shortest path algorithm and is able to select links with high transmissivity. This is because the shortest path algorithm only focuses on link distance and does not consider link quality. Furthermore, the proposed DQN algorithm is compared with both the shortest path algorithm and an LSTM-based RL algorithm in terms of the average fidelity of the end-to-end entanglement shared between communication parties. The implemented LSTM-based RL architecture comprises an LSTM layer with $64$ neurons, followed by a fully connected layer. This LSTM layer processes a state representation vector, which is then passed into the fully connected layer to generate Q-values. These Q-values represent the expected cumulative rewards for different actions. It is worth noting that the LSTM-based RL algorithm is trained on the dynamic network topology of SPARQ, similar to the proposed DQN algorithm. The corresponding results are illustrated in Fig. \ref{fig:DQN_vs_Shortest_EOF} for the comparison with the shortest path algorithm and in Fig. \ref{fig:DQN_vs_LSTM_EOF} for the comparison with the LSTM-based RL algorithm. In order to obtain these results, the communication parties are selected randomly from ground nodes. Then, the proposed DQN, the shortest path, and the LSTM-based RL algorithms are used to find the optimal routing path over $100$ satellite steps of satellites. Finally, the TPED policy is used to distribute on-demand entanglement between communication parties and the entanglement fidelity is measured. This experiment is repeated $100$ times and the average entanglement fidelity is recorded in each time step. 
As shown in Fig. \ref{fig:DQN_vs_Shortest_EOF}, the proposed DQN algorithm can distribute entangled pairs with the entanglement quality that is essential for successful teleportation (fidelity $>0.9$) most of the time, while the shortest path algorithm fails to meet this requirement most of the time. Specifically, the proposed DQN algorithm is able to meet this requirement $100\%$ of the time, while the shortest path algorithm is only able to meet this requirement $61\%$ of the time. This indicates that the proposed DQN algorithm improved the number of resolved teleportation requests by $39\%$ compared with the shortest path algorithm. This is because the shortest path algorithm prioritizes link distance over the link quality. Although both the proposed DQN and the LSTM-based RL algorithms follow the same pattern, the proposed DQN algorithm improves the end-to-end entanglement fidelity by about $2\%$ as shown in Fig. \ref{fig:DQN_vs_LSTM_EOF}. The comparable results can be attributed to the utilization of the same foundational RL algorithm, with both RL algorithms trained on the dynamic network topology of SPARQ. The observed improvement can be attributed to the effective utilization of the experience buffer mechanism, which plays a pivotal role in the DQN's outperformance of the LSTM-based RL. It is worth mentioning that implementing an experience buffer in the LSTM-based RL poses a significant challenge due to the inherent nature of LSTM, which necessitates the simultaneous input of multiple related time steps. Furthermore, we have calculated the confidence interval based on the normal distribution, and with $95\%$ confidence. We assert that the mean entanglement fidelity of the teleportation request solved by the proposed DQN algorithm falls within the range of $0.98$ to $0.99$.

\begin{figure}
    \centering
    \captionsetup{labelfont={color=black}}
    \includegraphics[width=3in]{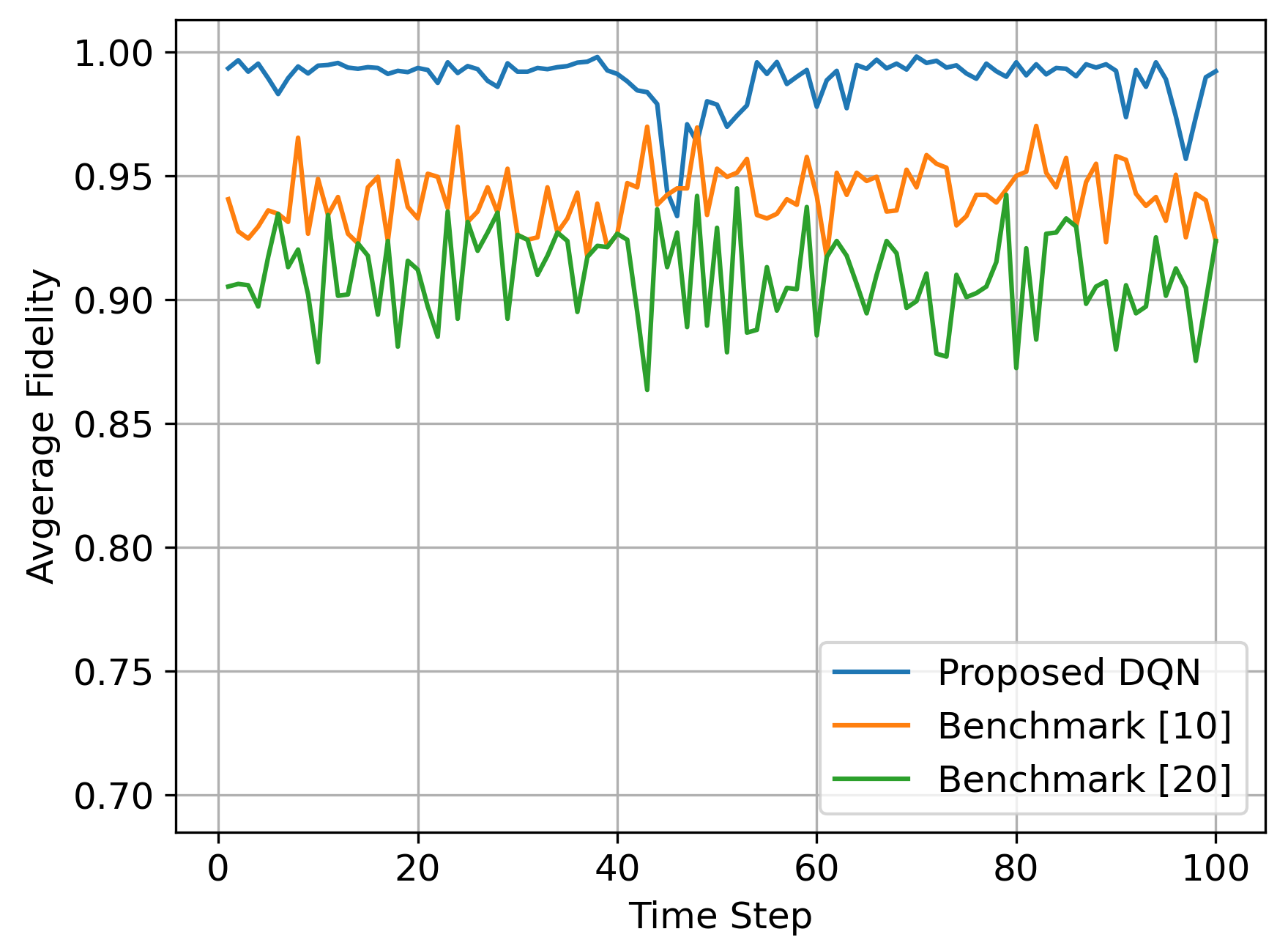}
    \caption{Comparing the proposed DQN algorithm with benchmarks \cite{24} and \cite{photonics11030268} in terms of average entanglement fidelity.}
    \label{fig:DQN_vs_Literature}
\end{figure}

Moreover, we compared the proposed DQN algorithm with two benchmarks proposed in \cite{24} and \cite{photonics11030268}. In this comparison, the communication parties are randomly selected from ground nodes, and each routing algorithm is executed to find the optimal path between them over $100$ time steps. Subsequently, the TPED policy is utilized to establish entanglement between the source and destination, and the resulting end-to-end entanglement fidelity is measured. This process is repeated $100$ times and the average entanglement fidelity is calculated. As shown in Fig. \ref{fig:DQN_vs_Literature}, the proposed DQN model outperforms both benchmarks. In particular, the DQN algorithm improves the average entanglement fidelity by about $6\%$ and $9\%$ compared with \cite{24} and \cite{photonics11030268}, respectively. The enhancement over the benchmark proposed in \cite{24} can be attributed to the consideration of both link quality and the number of hops in our DQN algorithm, whereas \cite{24} solely focuses on the number of hops. Similarly, the improvement over the benchmark proposed in \cite{photonics11030268} can be attributed to the joint consideration of link quality and the number of hops in finding the optimal path in our DQN algorithm, while \cite{photonics11030268} prioritizes the key rate.


\begin{figure}
    \centering
    \captionsetup{labelfont={color=black}}
    \includegraphics[width=3in]{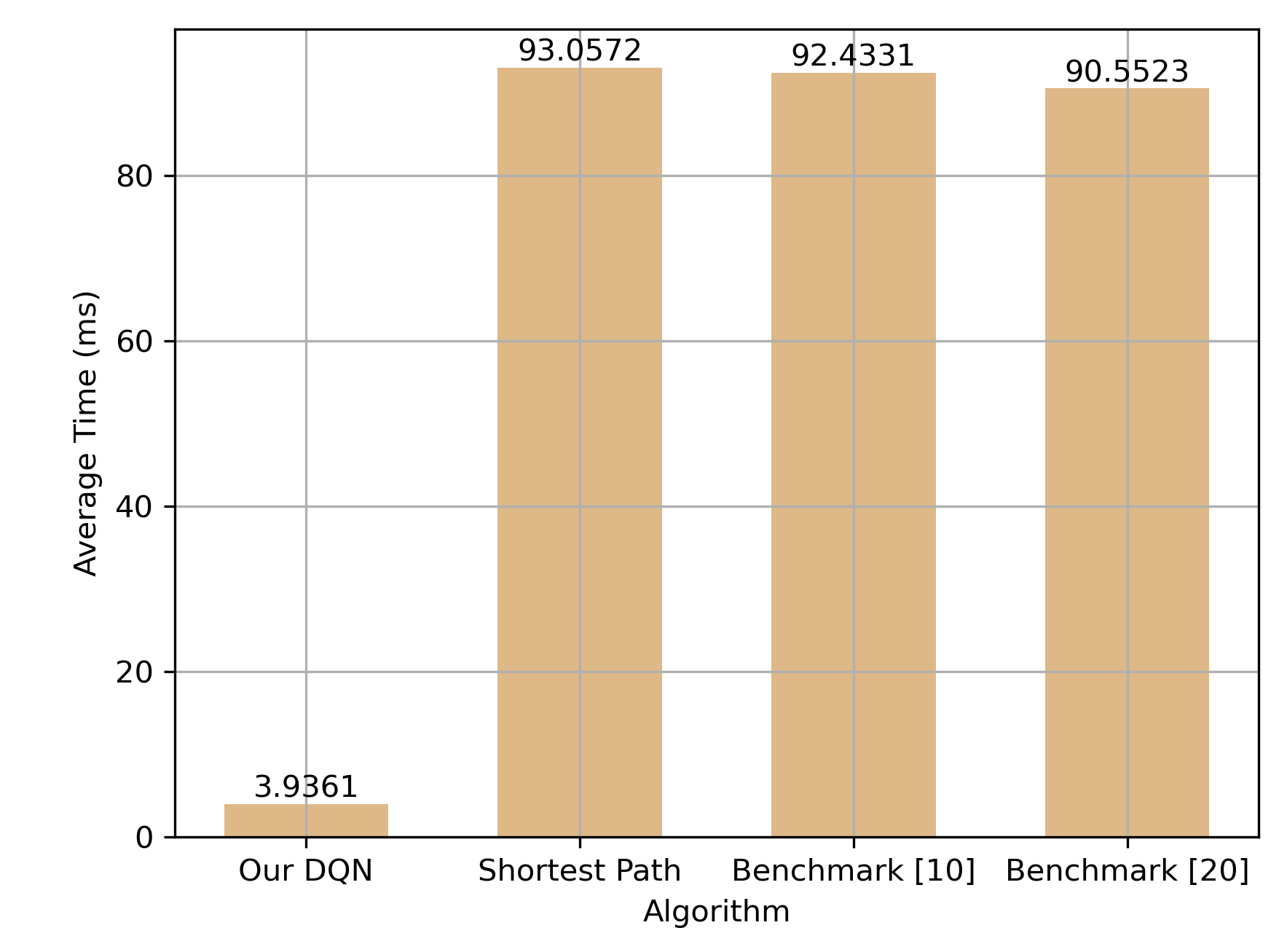}
    \caption{Comparing the proposed DQN algorithm with the shortest path algorithm and benchmark algorithms \cite{24} and \cite{photonics11030268} in terms of the average time required to find the optimal path between communication parties.}
    \label{fig:Time}
    \vspace{-3mm}
\end{figure}

In order to evaluate the overhead introduced by the proposed DQN algorithm, we conducted an experiment to compare the average time required for different algorithms to find the optimal path between a source and destination over $100$ trials. As shown in Fig. \ref{fig:Time}, the proposed DQN algorithm exhibits the lowest average time among all algorithms, with an average time of approximately $3.94$ milliseconds. This highlights the efficiency of our DQN algorithm in optimizing routing requests within SPARQ. In contrast, traditional methods such as the shortest path algorithm and the benchmarks \cite{24} and \cite{photonics11030268} require significantly more time, with their average times ranging from $90$ to $93$ milliseconds. This improvement can be attributed to the efficiency of RL, which simplifies the routing problem by learning through experience. As mentioned earlier, the training of the DQN agent was performed offline to ensure efficient real-time performance. Additionally, the efficient representation of the system state contributes to the agent's efficiency, as it requires only local knowledge to choose the next action. In contrast, benchmark algorithms require global knowledge, and they need to calculate the routing path for each communication request. The substantial reduction in simulation time achieved by our DQN algorithm underscores its practical viability for real-time routing optimization in dynamically changing network environments like SPARQ.



\begin{figure}
    \centering
    \captionsetup{labelfont={color=black}}
    \includegraphics[width=3in]{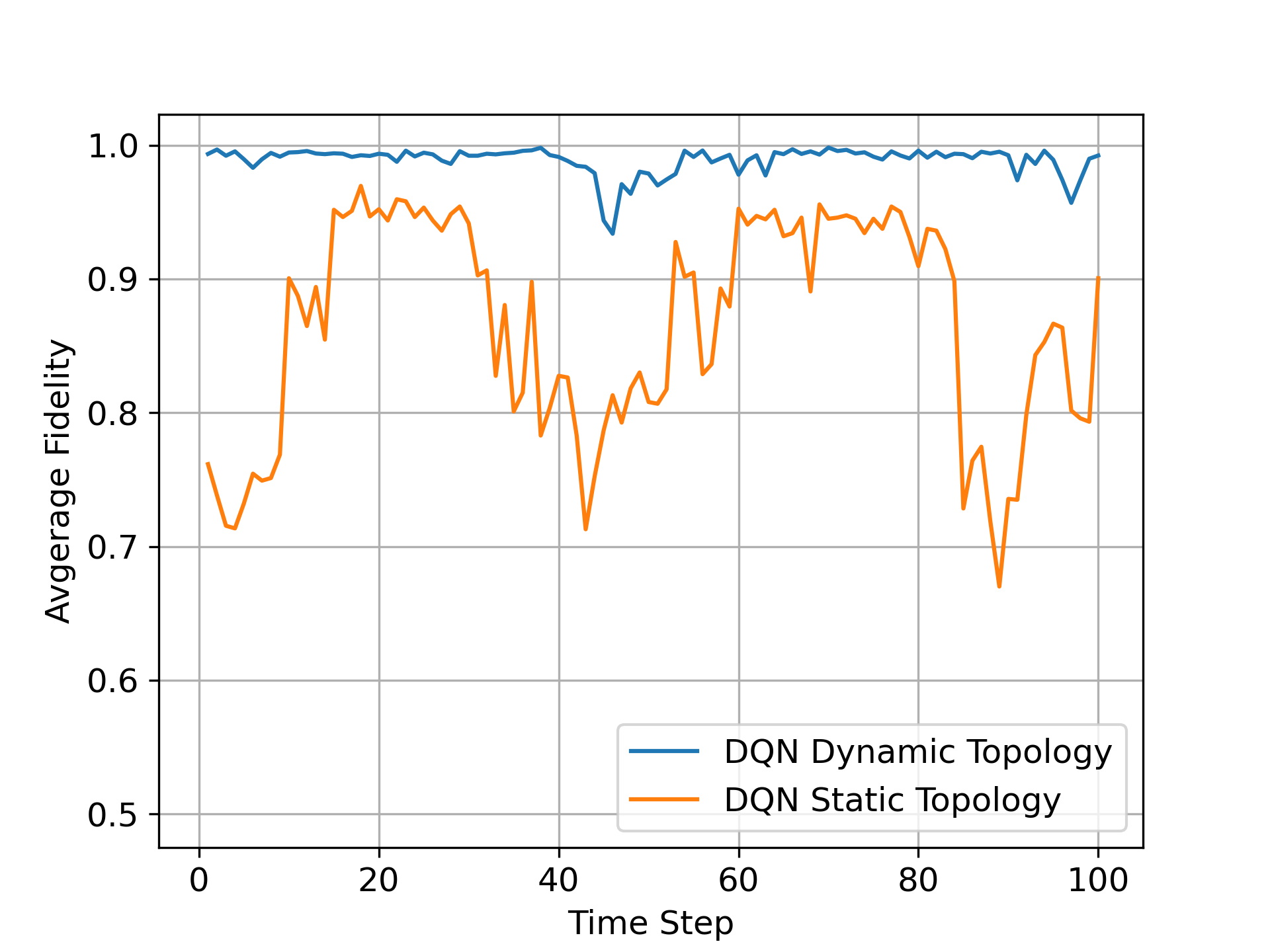}
    \caption{Comparing two versions of the DQN algorithm: one is trained on the dynamic network topology of SPARQ, while the other is trained on a static snapshot of SPARQ.}
    \label{fig:dynamic_static}
    \vspace{-3mm}
\end{figure}

Furthermore, the proposed DQN algorithm is compared with another DQN algorithm trained in the same setup on a single snapshot of the SPARQ network. This comparison aims to highlight the significance of training the agent on the dynamic network topology of SPARQ. Similarly, in this experiment, the communication parties are chosen randomly from ground nodes, and each DQN algorithm is applied to find the optimal path between the communication parties over $100$ time steps. Finally, the TPED policy is used to establish entanglement between the source and destination and the entanglement fidelity is measured and recorded. This experiment is repeated $100$ times and the average entanglement fidelity is calculated. As shown in Fig. \ref{fig:dynamic_static}, the DQN algorithm that is trained on the dynamic network topology outperforms the other algorithm trained on a static snapshot of the SPARQ network. Specifically, the proposed DQN algorithm achieves an average entanglement fidelity of $0.988$, while the other algorithm achieves an average entanglement fidelity of $0.86$, which indicates an improvement by about $15\%$. This is because training the agent on the dynamic network topology allows the agent to capture all states of the network graph and choose the optimal route based on the encountered network state.

\begin{figure}
    \centering
    \captionsetup{labelfont={color=black}}
    \includegraphics[width=3in]{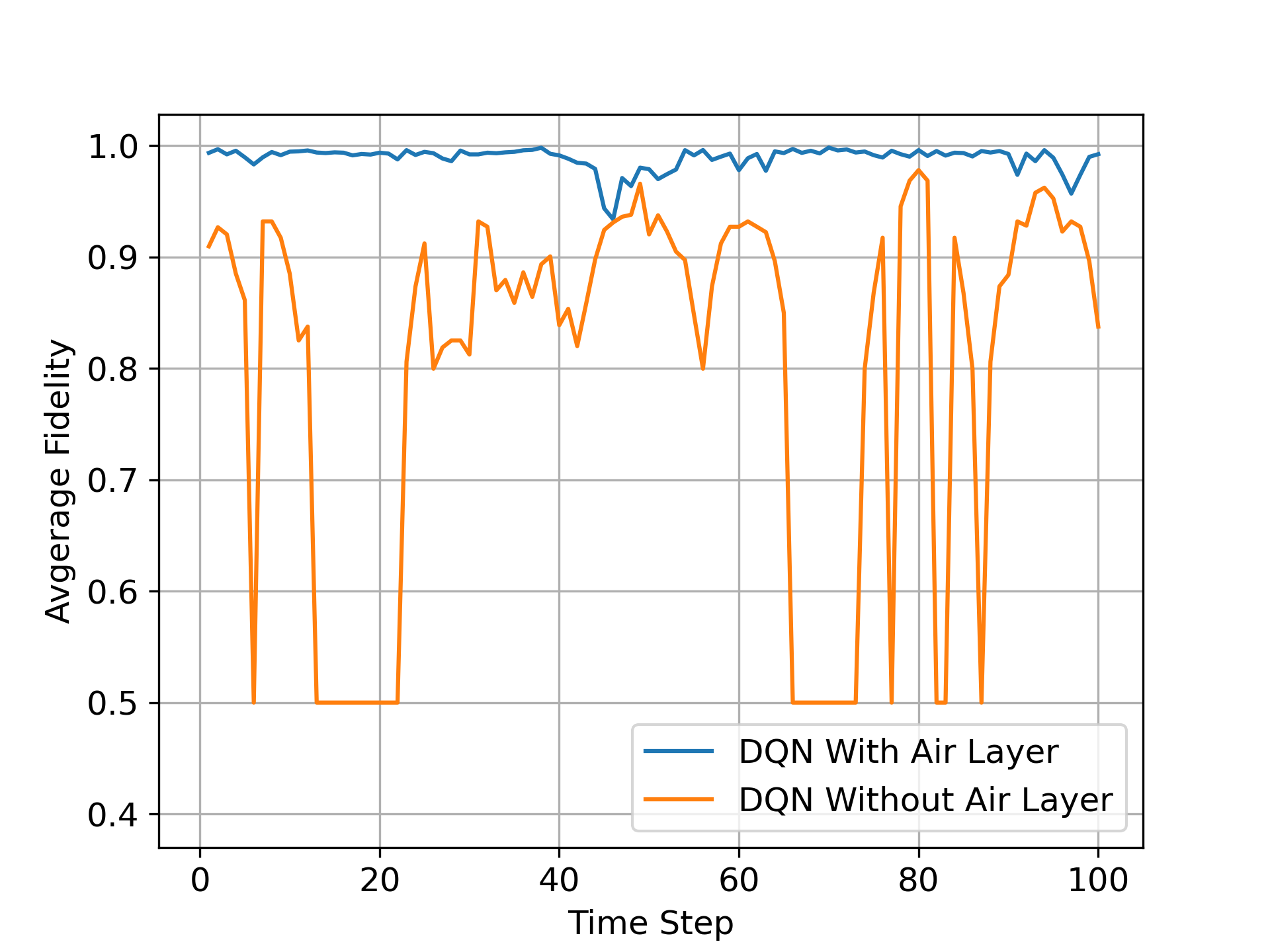}
    \vspace{-1mm}
    \caption{Comparing the SPARQ network performance with a network that spans only space and ground layers.}
    \label{fig:DQN_no_uavs}
    \vspace{-3mm}
\end{figure}

In order to highlight the importance of the air layer within SPARQ, we trained another DQN algorithm to find the optimal routing path without the existence of the air layer. In other words, the DQN algorithm is trained on a network that spans only the space and the ground layers. Similarly, The source and the destination are selected randomly from the ground layer. 
To extend satellite coverage in the absence of HAPs and facilitate the establishment of paths between communication parties, we adjusted the transmissivity threshold to $0.5$. Each DQN algorithm is used to find the optimal routing path within the corresponding network over $100$ time steps. Then, the TPED policy is used to establish entanglement between the source and the destination, and the entanglement fidelity is measured. The average is calculated across $100$ different trials and the results are presented in Fig. \ref{fig:DQN_no_uavs}. The proposed DQN algorithm in SPARQ achieves an average entanglement fidelity of $0.988$, while the architecture that spans only satellite and ground layers achieves an average entanglement fidelity of $0.8$, which indicates an improvement by $23.5\%$. As shown in Fig. \ref{fig:DQN_no_uavs}, the performance of the SPARQ network with the existence of the air layer is much better than the performance without the air layer. This is because of two reasons. Firstly, the air layers bridge the considerable distance between satellite nodes and ground nodes. Secondly, the air layer satisfies teleportation requests when satellites are out of range.

\subsection{Discussion}
\textcolor{black}{The proposed SPARQ network and its associated algorithms are evaluated across several key performance metrics, including average entanglement fidelity, the number of resolved teleportation requests, memory consumption, and the time required to find a routing path between communication parties. For instance, an experiment comparing SPARQ with a network that spans only space and ground layers highlights the critical role of the air layer. The inclusion of the air layer in SPARQ achieves a $23.5\%$ improvement in average entanglement fidelity, underscoring its significant contribution to overall network performance. Within the SPARQ architecture, the TPED policy improves average entanglement fidelity by $3\%$ compared to the intuitive approach. The proposed DQN algorithm improves the fidelity compared to state-of-the-art benchmarks while yielding significant gains in other performance metrics. For instance, the proposed TPED policy achieves a $50\%$ reduction in memory consumption, underscoring its efficiency in managing quantum resources. Additionally, the proposed DQN algorithm enhances the number of resolved teleportation requests by $39\%$ compared to the shortest path baseline. Furthermore, the proposed DQN algorithm also significantly reduces the time required to find the routing path, demonstrating a substantial improvement in speed over state-of-the-art algorithms, thereby optimizing network routing and resource utilization.}

\section{Conclusion}
\label{sec:Conclusion}
In this paper, we have proposed a space-air-ground quantum network designed to enable quantum Internet applications in the near term. SPARQ aims to allow on-demand and seamless entanglement distribution, which enables applications such as distributed quantum computing, quantum communications, quantum sensing, and quantum security. 
We have then developed a DQN-based reinforcement learning algorithm to efficiently solve the routing problem between communication parties within SPARQ. Additionally, we have proposed an entanglement distribution policy in order to establish on-demand entanglement between communication parties. Furthermore, we have upgraded an existing quantum network simulator to carry out SPARQ's simulation and evaluate the performance of SPARQ, the proposed DQN algorithm, and the entanglement distribution policies. 
Simulation results show that the proposed TPED policy improved the average end-to-end entanglement fidelity by $3\%$ and reduced the memory consumption by $50\%$ compared with the intuitive approach. The proposed DQN algorithm is compared with the shortest path algorithm, an LSTM-based RL algorithm, two state-of-the-art benchmark algorithms, and another DQN algorithm that is trained on a static snapshot of SPARQ. The proposed DQN algorithm demonstrated significant improvements across various algorithms. Compared with the shortest path algorithm, it enhanced the number of resolved teleportation requests by $39\%$. Additionally, in comparison with an LSTM-based algorithm, it achieved a $2\%$ increase in average entanglement fidelity. \textcolor{black}{Furthermore, when benchmarked against two state-of-the-art algorithms, the proposed DQN model improved the average fidelity by $6\%$ and $9\%$, respectively.} Moreover, it showcased an average improvement of $15\%$ in end-to-end entanglement fidelity compared to a DQN trained on a snapshot of SPARQ. Also, to highlight the importance of the air layer, we have conducted an experiment to compare the efficiency of SPARQ and the network spanning only space and ground layers. The results also show that communication within SPARQ is more efficient when the air layer is included. SPARQ enhanced the average end-to-end entanglement fidelity by $23.5\%$ compared with the network that spans only space and ground layers. The attained entanglement fidelity has the potential for improvement through entanglement purification but it will consume additional quantum resources. However, we leave the exploration of efficient entanglement purification to future work. Additionally, future research may consider identifying optimal satellite trajectories to improve coverage and connectivity between distant nodes. \textcolor{black}{Moreover, future extensions could improve the SPARQ simulator by incorporating more realistic physical layer models that address hardware imperfections, such as quantum memory decoherence and deficiencies in quantum devices like repeaters and detectors.}

\begin{appendices}
\section{Simulation Parameters}
\label{app:sim}
\textcolor{black}{
The attenuation coefficient is set to $0.15$ dB/km for fiber optical channels as it is used in \cite{106}. 
The simulation parameters for the FSO channels, shown in Table \ref{tab:parameters}, are adjusted based on \cite{107}. 
The learning rate $\beta$ is set to $0.001$, while the exploration rate ($\epsilon$) is initialized to be $1$ and decays by $0.9995$ until reaches $0.01$. The Adam optimizer is used to minimize the loss by updating the weights of the neural network. The average training loss is visualized in Fig. \ref{fig:loss}, where the average is calculated based on all losses encountered during the $100$ mini-episodes used to train the model on the current network graph. The simulation parameters are summarized in Table \ref{tab:parameters}.}
\begin{table}[H]
    \centering
    \renewcommand{\arraystretch}{1.3}
    \caption{Simulation parameters.}
    \label{tab:parameters}
    \begin{tabular}{|c|c|}
        \hline
        \textbf{Parameter}&\textbf{Value}\\\hline
         Number of satellites&$20$\\
         Altitude of satellites&$500$ km\\
         Number of HAPs&$10$\\
         Flying height of HAPs&$50$ km\\
         Number of ground nodes&$24$\\
         Number of episodes&$500$\\
         Number of mini-episodes&$100$\\
         $\beta$&$0.001$\\
         $\epsilon$& $1-0.01$\\
         decay rate of $\epsilon$&$0.9995$\\
         $\gamma$&$0.99$\\
         Optimizer& Adam\\
         Hidden layers&$2$\\
         Number of neurons& $64, 128$\\
         Activation function& ReLU\\
         Transmissivity Threshold& $0.7$\\
         $\alpha$&0.15 dB/km\\
         $\eta_{\mbox{\tiny\itshape eff}}$&1\\
         $a_{\mbox{\tiny\itshape R}}$& $5$ m \\
         $R_0$& $\infty$\\
         $\alpha_0(\lambda)$& $5\times10^{-6}$m$^{-1}$\\
         $l_0$&$1$ mm\\
         $w_0$&$20$ cm\\
         $\lambda$&$800$ nm\\\hline
    \end{tabular}
\end{table}

\section{Training Dataset}
\label{app:dataset}
\textcolor{black}{Our dataset is dynamically generated during the training phase as the DQN agent interacts directly with the network environment to learn from experience. To facilitate the learning process, we simulate the SPARQ network before the actual system is operational, allowing the agent to encounter different network graphs and learn optimal routing policy. Transmissivities to adjacent nodes serve as a feature for training and identifying the current network state. For each graph, routing requests are randomly generated between ground nodes, and the agent's routing decisions lead to either rewards or penalties. Starting from the source node, the agent selects actions to visit one of the adjacent nodes, and it continues until reaching the destination. Initially, the agent takes random actions to explore diverse paths and strategies for finding the optimal route for entanglement distribution. As training progresses, the DQN agent uses the received rewards to update its internal Q-values. The Q-values represent the expected cumulative reward for taking a particular action in a given state. Throughout training, the exploration rate $(\epsilon)$ gradually decreases, leading to a reduced dependence on random actions. Instead, the agent increasingly prioritizes actions based on its learned policy. This transition from exploration to exploitation occurs until the agent converges toward a strategy for finding an effective path for entanglement distribution. Here, we note that the agent relies solely on local information to determine the next action at each node. This local information, represented by transmissivities to adjacent nodes, provides insight into channel quality and connectivity. Since satellite trajectories are relative and periodic, this local information effectively represents the network state, enabling the agent to optimize routing requests based on encountered network states.}

\textcolor{black}{Several key factors underscore the significance of this dataset. Firstly, it enables the agent to interact with diverse network topologies and configurations, reflecting the dynamic nature of SPARQ. This interaction with a wide range of network scenarios ensures that the trained model can effectively adapt to varying environmental conditions. Secondly, it allows the agent to optimize routing requests between several communication nodes at each network state during training. Consequently, the dataset facilitates the development of a comprehensive policy for identifying the optimal routing path for entanglement distribution.}

\section{Nodes Configurations}
\label{app:config}
\textcolor{black}{The coordinates for the ground nodes are shown in Table \ref{tab:ground}, and the coordinates for the HAPs are listed in Table \ref{tab:HAPs}. Furthermore, the orbital configurations of the satellites are detailed in Table \ref{tab:satellites}, including their inclination, right ascension of the ascending node (RAAN), and altitude. Notably, these orbital configurations can be used in the STK simulator to replicate the trajectories of the satellites.}
\begin{table}[H]
    \renewcommand{\arraystretch}{1.5}
    \centering
    \captionsetup{labelfont={color=black}}
    \caption{\textcolor{black}{Coordinates of ground nodes.}}
    {\color{black}\begin{tabular}{|c|c|c|}
    \hline
    \multicolumn{3}{|c|}{Ground Nodes}\\
    \hline

($34.315$,$112.732$)&($34.336$,$112.736$)&($34.357$,$112.740$)\\
($34.378$,$112.746$)&($34.399$,$112.750$)&($34.415$,$112.754$)\\
($72.823$,$84.759$)&($72.844$,$84.765$)&($72.865$,$84.771$)\\
($72.876$,$84.777$)&($-0.062$,$-60.266$)&($-0.083$,$-60.270$)\\
($-0.103$,$-60.278$)&($-0.113$,$-60.285$)&($-0.126$,$33.812$)\\
($-0.105$,$33.816$)&($59.561$,$102.211$)&($59.568$,$102.251$)\\
($22.455$,$90.777$)&($22.434$,$90.774$)&($34.411$,$66.991$)\\
($34.426$,$66.972$)&($60.064$,$77.805$)&($60.043$,$77.813$)

     \\\hline
    \end{tabular}}
    \label{tab:ground}
\end{table}

\begin{table}[H]
    \renewcommand{\arraystretch}{1.5}
    \centering
    \captionsetup{labelfont={color=black}}
    \caption{\textcolor{black}{Coordinates of HAPs.}}
    {\color{black}\begin{tabular}{|c|c|c|}
    \hline
    \multicolumn{3}{|c|}{HAPs}\\
    \hline
($34.368$,$112.738$)&($-0.062$,$-60.266$)&($-0.162$,$-60.235$)\\
($-0.105$,$33.832$)&($36.644$,$113.444$)&($34.418$,$66.979$)\\
($60.044$,$77.848$)&($59.561$,$102.251$)&($75.834$,$84.763$)\\
($22.485$,$90.777$)&&

     \\\hline
    \end{tabular}}
    \label{tab:HAPs}
\end{table}

\begin{table}[H]
    \renewcommand{\arraystretch}{1.5}
    \centering
    \captionsetup{labelfont={color=black}}
    \caption{\textcolor{black}{Satellites orbital configurations.}}
    {\color{black}\begin{tabular}{|c|c|c|}
    \hline
    \multicolumn{3}{|c|}{Satellites}\\
    \hline
    Inclination (deg)& RAAN (deg)& Altitude (km)\\\hline
     $36$&$126$&$500$\\
     $54$&$126$&$500$\\
     $105$&$126$&$500$\\
     $36$&$233$&$500$\\
     $72$&$304$&$500$\\
     $108$&$15$&$500$\\
     $125$&$53$&$500$\\
     $144$&$91$&$500$\\
     $0$&$0$&$500$\\
     $18$&$158$&$500$\\
     $73$&$158$&$500$\\
     $90$&$160$&$500$\\
     $127$&$160$&$500$\\
     $144$&$160$&$500$\\
     $163$&$160$&$500$\\
     $0$&$200$&$500$\\
     $18$&$228$&$500$\\
     $55$&$300$&$500$\\
     $90$&$13$&$500$\\
     $162$&$160$&$500$\\
     \hline
    \end{tabular}}
    \label{tab:satellites}
\end{table}

\end{appendices}

\bibliographystyle{IEEEtran}
\bibliography{References}
\vspace{-10mm}
\begin{IEEEbiography}[{\includegraphics[width=1in,height=1.25in,clip,keepaspectratio]{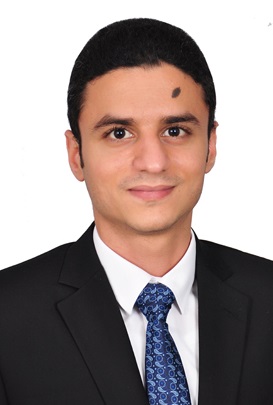}}]{Mohamed Shaban}(mmibrahims42@tntech.edu) is a Ph.D. student with the Computer Science Department, Tennessee Technological University. He is a Teaching Assistant with the Mathematics Department, Faculty of Education, Alexandria University, Egypt. He obtained his B.Sc. degree in Computer Science and M.Sc. degree in quantum computing, both from the Faculty of Science, Alexandria University, Egypt. 
\end{IEEEbiography}

\vspace{-12mm}
\begin{IEEEbiography}[{\includegraphics[width=1in,height=1.25in,clip,keepaspectratio]{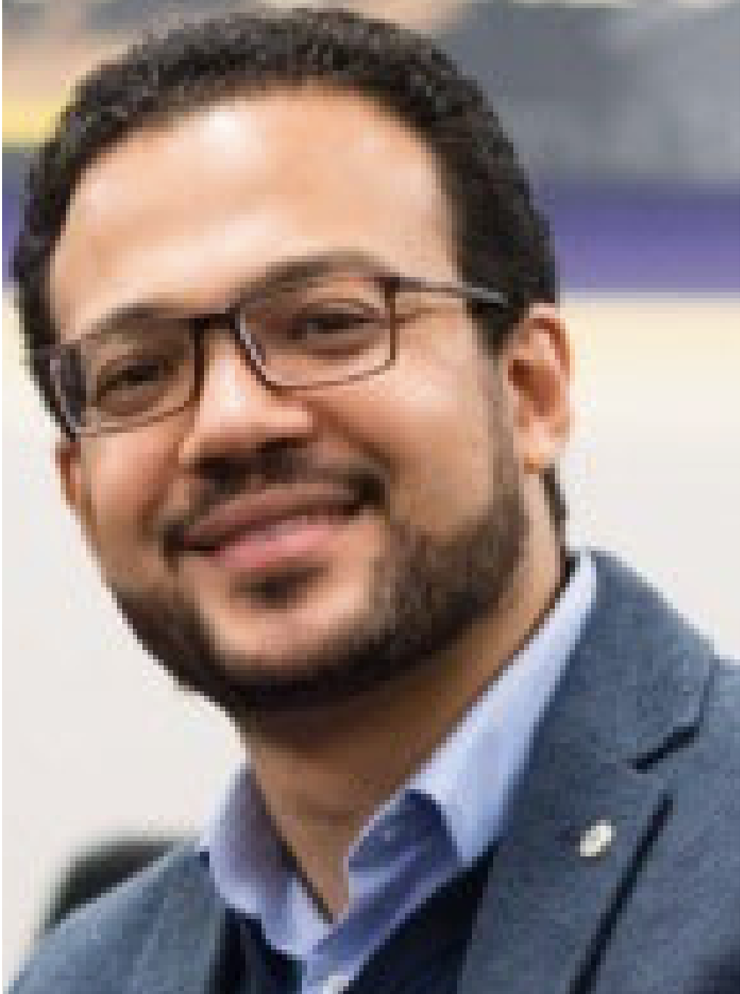}}]{Muhammad Ismail} (S'10-M'13-SM'17) 
received the B.Sc. (Hons.) and M.Sc. degrees in electrical engineering (electronics and communications) from Ain Shams University, Cairo, Egypt, in 2007 and 2009, respectively, and the Ph.D. degree in electrical and computer engineering from the University of Waterloo, Waterloo, ON, Canada, in 2013. He is the Director of the Cybersecurity Education, Research, and Outreach Center (CEROC) and an Associate Professor with the Department of Computer Science, Tennessee Technological University, Cookeville, TN, USA. He was a co-recipient of the best paper awards in the IEEE ICC 2014, the IEEE GLOBECOM 2014, the SGRE 2015 and 2024, the Green 2016, the IEEE IS 2020, and the Best Conference Paper Award from the IEEE Communications Society Technical Committee on Green Communications and Networking for his publication in IEEE ICC 2019. He is a Track chair in the IEEE Globecom 2024. He was a Track Co-Chair in the IEEE SmartGridComm 2023, Workshop Co-Chair of the IEEE Greencom 2018, Track Co-Chair of the IEEE VTC 2017 and 2016, the Publicity and Publication Co-Chair of the CROWNCOM 2015, and the Web-Chair of the IEEE INFOCOM 2014. He organized workshops on topics of quantum cryptography and quantum machine learning in IEEE QCE 2024, ICMLA 2024, and WiCyS 2024. He was an Associate Editor of the IET Communications, PHYCOM, and the IEEE Transactions on Green Communications and Networking. He was an Editorial Assistant of the IEEE Transactions on Vehicular Technology, from 2011 to 2013. He is an Associate Editor of the IEEE Internet of Things Journal and the IEEE Transactions on Vehicular Technology. He has been a technical reviewer of several IEEE conferences and journals.
\end{IEEEbiography}

\begin{IEEEbiography}[{\includegraphics[width=1in,height=1.25in,clip,keepaspectratio]{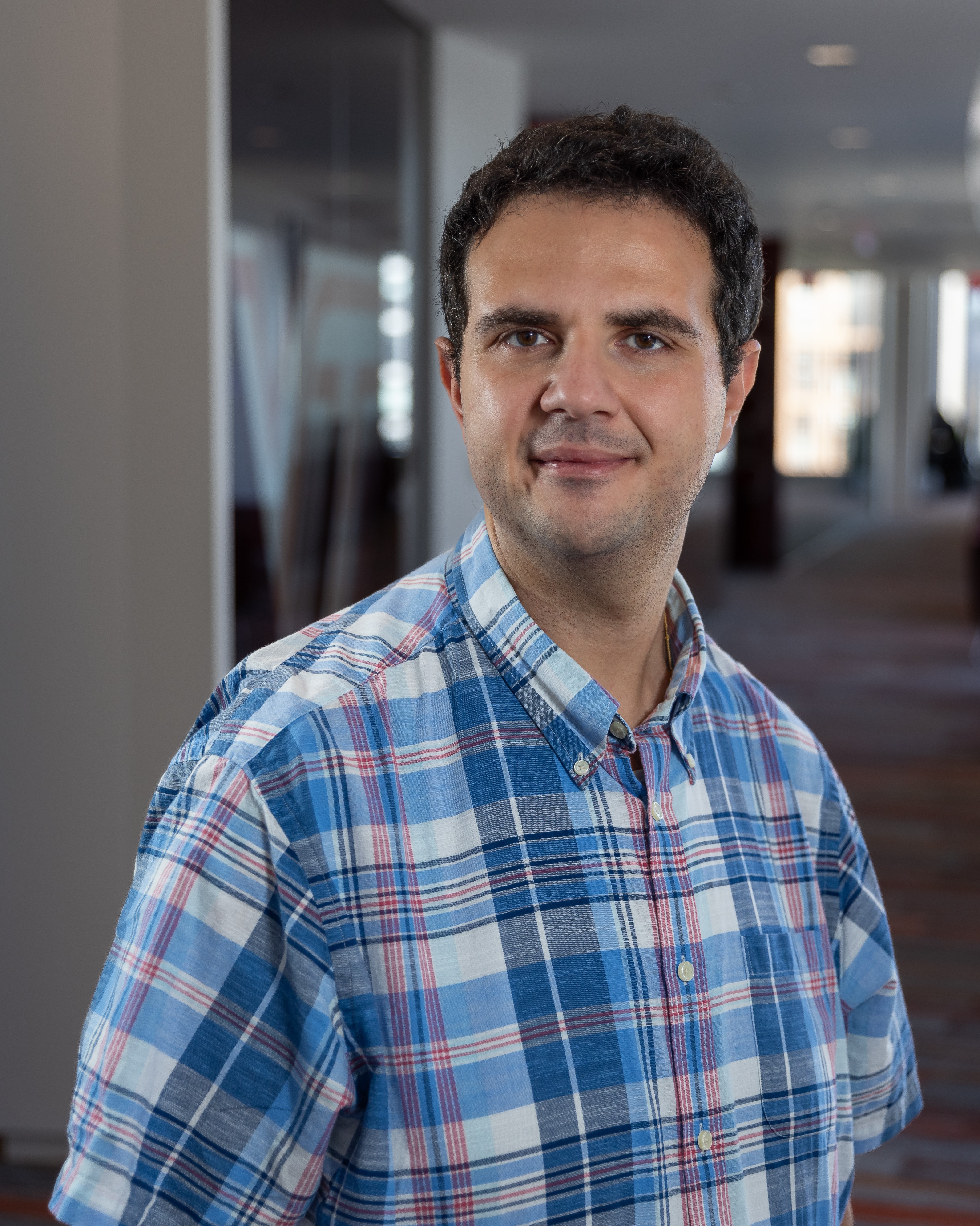}}]{Walid Saad} (S’07, M’10, SM’15, F’19) 
received his Ph.D degree from the University of Oslo, Norway in 2010. He is currently a Professor at the Department of Electrical and Computer Engineering at Virginia Tech, where he leads the Network intelligEnce, Wireless, and Security (NEWS) laboratory. His research interests include wireless networks (5G/6G/beyond), machine learning, game theory, quantum communications/learning, security, UAVs, semantic communications, cyber-physical systems, and network science. Dr. Saad is a Fellow of the IEEE. He is also the recipient of the NSF CAREER award in 2013, the AFOSR summer faculty fellowship in 2014, and the Young Investigator Award from the Office of Naval Research (ONR) in 2015. He was the (co-)author of twelve conference best paper awards at IEEE WiOpt in 2009, ICIMP in 2010, IEEE WCNC in 2012, IEEE PIMRC in 2015, IEEE SmartGridComm in 2015, EuCNC in 2017, IEEE GLOBECOM (2018 and 2020), IFIP NTMS in 2019, IEEE ICC (2020 and 2022), and IEEE QCE in 2023. He is the recipient of the 2015 and 2022 Fred W. Ellersick Prize from the IEEE Communications Society,  of the IEEE Communications Society Marconi Prize Award in 2023, and of the IEEE Communications Society Award for Advances in Communication in 2023. He was also a co-author of the papers that received the IEEE Communications Society Young Author Best Paper award in 2019, 2021, and 2023. Other recognitions include the 2017 IEEE ComSoc Best Young Professional in Academia award, the 2018 IEEE ComSoc Radio Communications Committee Early Achievement Award, and the 2019 IEEE ComSoc Communication Theory Technical Committee Early Achievement Award. From 2015-2017, Dr. Saad was named the Stephen O. Lane Junior Faculty Fellow at Virginia Tech and, in 2017, he was named College of Engineering Faculty Fellow. He received the Dean's award for Research Excellence from Virginia Tech in 2019. He was also an IEEE Distinguished Lecturer in 2019-2020.  He has been annually listed in the Clarivate Web of Science Highly Cited Researcher List since 2019. He currently serves as an Area Editor for the IEEE Transactions on Communications. He is the Editor-in-Chief for the IEEE Transactions on Machine Learning in  Communications and Networking.
\end{IEEEbiography}

\vfill

\EOD
\end{document}